\documentclass[aps,pra,onecolumn,notitlepage,10pt,amsmath,amssymb]{revtex4-2}

\usepackage[english, bidi=default, provide=*]{babel}
\babelprovide[import, onchar=ids fonts]{english}

\usepackage{amsthm}
\usepackage{booktabs}
\usepackage{xcolor}

\usepackage{graphicx}

\newtheorem{proposition}{Proposition}

\begin{document}

\title{Invariant Phase-Space Boundaries Across Coulomb Cores and Tunneling Barriers}

\author{Carlos A. Arango}
\affiliation{Department of Pharmaceutical and Chemical Sciences, Universidad Icesi, Cali, Colombia, 760031}

\date{\today}

\begin{abstract}
Invariant phase-space boundaries and contracting bundles governing stationary Schr\"odinger flow across singular Coulomb cores and classically forbidden tunneling barriers in one dimension are investigated. Grounded in Pr\"ufer and quasi-derivative spectral theory, trajectories in unreduced phase space evolve directly in real coordinates without coordinate singularities, complex coordinate scaling, or artificial Dirichlet box walls. In classically forbidden regions, the physical decaying solution acts as an invariant ordering boundary on the autonomous Pr\"ufer cylinder, allowing opposite nullcline exits to certify bracketing of the decaying trajectory without prior knowledge of the exact solution. Building on this geometric structure, \emph{Generational Stable-Manifold Bracketing} (GSMB) pairs Kato cusp initialization at the core with periodic transverse chord re-bracketing across adaptive handoff radii, backward stability inversion for full-domain wavefunction reconstruction, and two-sided real Wronskian matching at an interior barrier plane to extract bound states, real scattering phase shifts, and Wigner time delays. Benchmarks on radial atomic hydrogen and an $\ell = 2$ screened-Coulomb shape resonance illustrate contraction across forbidden regions, exact core cusp launching, and agreement with Siegert and finite-element reference calculations.
\end{abstract}

\keywords{Invariant phase-space boundaries; Pr\"ufer transformation; contracting manifolds; quantum tunneling; shape resonances; Kato cusp condition; Wigner time delay; radial Schr\"odinger equation}

\maketitle

\section{Introduction}
\label{sec:intro}

Accurate propagation of stationary wavefunctions across singular atomic potentials and classically forbidden tunneling barriers remains a foundational problem in atomic, molecular, and scattering physics. For decades, direct numerical integration of one-dimensional (1D) Schr\"odinger equations has relied primarily on Numerov's fourth-order finite-difference recurrence~\cite{Numerov1924,Cooley1961} or on logarithmic-derivative propagation schemes such as Johnson's renormalized Numerov algorithm~\cite{Johnson1977} and Manolopoulos's log-derivative integrator~\cite{Manolopoulos1986}. While effective in standard bound-state and elastic scattering calculations, these methods face structural obstacles near attractive Coulomb cores and narrow tunneling resonances.

Direct forward integration via Numerov's recurrence fails rapidly in classically forbidden regions. In the phase plane spanned by $(\psi, \psi')$, a forbidden barrier is governed by a non-autonomous hyperbolic saddle flow with contracting and expanding streamlines. Because the local vector field rotates continuously with position at fixed energy, finite step-size discretization lag and floating-point roundoff inevitably push a single trajectory across the separatrix into exponential growth before the asymptotic tail is resolved. Bidirectional shooting~\cite{Cooley1961} mitigates single-trajectory overflow, while logarithmic-derivative propagators~\cite{Johnson1977,Manolopoulos1986} suppress overflow by evolving nonlinear Riccati ratios, but at the cost of pole singularities at every wavefunction node and loss of direct trajectory geometry.

Matrix diagonalization on finite grids or basis sets---often paired with complex coordinate scaling~\cite{Aguilar1971,Balslev1971,Moiseyev2011} or complex absorbing potentials~\cite{Seideman1992,Riss1993} for resonances---requires Dirichlet boundary conditions that force the wavefunction to vanish at inner and outer cutoffs. Near an attractive nuclear singularity, Dirichlet regularization distorts Kato's cusp condition~\cite{Kato1957} (especially in smooth basis or pseudospectral expansions), while an outer wall discretizes open continua into artificial box states.

Molano and Arango~\cite{Molano2021} introduced a geometric alternative by formulating stationary propagation as a linear flow in the unreduced phase space $(\psi, \psi')$. Evolving polar angle and amplitude simultaneously avoids Riccati node blowups, preserves Wronskians and normalization, and accommodates exact Kato cusp initialization natively. Generalized Pr\"ufer spectral theory for Sturm--Liouville equations with quasi-derivative terms~\cite{Bandyopadhyay2025} further establishes bijective equivalence with the wave equation, nodal regularity, and strict energy monotonicity even with radial geometric terms $g(r) = 2/r$, while Solarte \emph{et al.}~\cite{Solarte2025} demonstrated the framework on piecewise-constant wells and barriers. Yet on spatially varying potentials, hyperbolic instability still causes finite-step errors to push trajectories from collapsing into exploding orbits across forbidden barriers.

The central purpose of this work is not merely to introduce another numerical propagation scheme, but to identify the geometric phase-space structure underlying why direct propagation fails across classically forbidden barriers and how that failure is controlled. Specifically, exponential contamination in non-autonomous Schr\"odinger flow arises from three coupled mechanisms: local hyperbolic saddle amplification, spatial eigenframe rotation, and continuous unstable-mode injection. On the autonomous Pr\"ufer cylinder, however, the physical decaying solution forms an impenetrable invariant ordering boundary inside a collapsing stream corridor, allowing opposite nullcline exits to certify bracketing without prior knowledge of the exact wavefunction. Building directly on this topological and covariant structure, \emph{Generational Stable-Manifold Bracketing} (GSMB) is formulated as a purely real phase-space propagation framework pairing energy-independent Kato cusp initialization with periodic transverse chord re-bracketing to confine orbits within the contracting corridor, retroactive backward stability inversion for full-domain wavefunction reconstruction, and two-sided real Wronskian matching at an interior barrier plane without artificial box walls. Evaluating these matching conditions and real scattering phase shifts yields bound eigenvalues, eigenfunctions, and Wigner time delays~\cite{Wigner1955,Smith1960} directly on the real energy axis.

This geometric structure is examined across two complementary benchmarks: unreduced radial atomic hydrogen ($1s$, $2s$, and $15s$ across nearly seven orders of magnitude in radius) and an $\ell = 2$ screened-Coulomb centrifugal barrier supporting a narrow shape resonance, compared against analytical solutions, complex-energy Siegert references, and finite-element cross-checks. Launching directly from the exact Kato core ray avoids near-origin mesh dependence; hydrogen isolates one-sided invariant ordering, escape-horizon suppression, and backward stability inversion across semi-infinite tails, whereas the screened-Coulomb resonance tests two-sided Wronskian barrier matching, phase-shift dispersion, and real-axis local density of states (LDOS) profiling directly on the differential equation.

Section~\ref{sec:theory} presents the Pr\"ufer and covariant-frame theory, Kato core initialization, generational bracketing and backward stability inversion, and real-axis Wronskian/time-delay matching. Section~\ref{sec:results} reports the hydrogen and screened-Coulomb resonance benchmarks, and Section~\ref{sec:conclusions} discusses broader comparisons and practical guidelines.

\section{Theory and Methods}
\label{sec:theory}

\subsection{Geometric Phase-Space Flow and Generalized Pr\"ufer Equivalence}

In atomic units ($\hbar = m_e = e = 1$), a stationary quantum state at energy $E$ is governed along a continuous spatial coordinate $r \in [r_{\min}, r_{\max}]$ by the one-dimensional Schr\"odinger equation
\begin{equation}
\frac{d^2\psi(r)}{dr^2} + g(r)\frac{d\psi(r)}{dr} + k^2(r; E)\,\psi(r) = 0,
\label{eq:radial_schrodinger}
\end{equation}
where $k^2(r; E) \equiv 2\mu\bigl[E-V_{\mathrm{eff}}(r)\bigr]$ is the local squared wavenumber for reduced mass $\mu$ and effective potential $V_{\mathrm{eff}}(r) = V(r) + \ell(\ell+1)/(2\mu r^2)$. The geometric damping coefficient equals $g(r) = 2/r$ in unreduced spherical radial coordinates and $g(r) = 0$ in one-dimensional Cartesian coordinates (with $\ell = 0$). Following the geometric phase-space formulation of Molano and Arango~\cite{Molano2021}, the derivative coordinate $p(r) \equiv \psi'(r)$ is introduced to form the unreduced phase-space state vector $\mathbf{x}(r) = (\psi(r), p(r))^\top \in \mathbb{R}^2$. Equation~\eqref{eq:radial_schrodinger} then becomes the linear non-autonomous dynamical system
\begin{equation}
\frac{d\mathbf{x}(r)}{dr} = \mathbf{A}(r; E)\mathbf{x}(r), \qquad
\mathbf{A}(r; E) = \begin{pmatrix}
0 & 1 \\
-k^2(r; E) & -g(r)
\end{pmatrix}.
\label{eq:phase_ode}
\end{equation}
In polar phase-space coordinates, any nontrivial trajectory $\mathbf{x}(r) = R(r)(\cos\theta(r), \sin\theta(r))^\top$ is parameterized by a strictly positive Euclidean radius $R(r) = \sqrt{\psi^2(r) + p^2(r)} > 0$ and a projective ray angle $\theta(r)$ satisfying
\begin{equation} 
\tan\theta(r)=\frac{p(r)}{\psi(r)}, \qquad \frac{p(r)}{\psi(r)} = \frac{d}{dr}\ln|\psi(r)| \quad\text{for }\psi(r)\neq0. 
\label{eq:ray_angle} 
\end{equation}

Geometrically, the ray angle $\theta(r)$ is a projective representation of the generalized Pr\"ufer phase~\cite{Prufer1926,Atkinson1964,Zettl2005,Johnson1978,Bandyopadhyay2025}, which has long served in chemical physics to establish node-counting and spectral-monotonicity theorems for Schr\"odinger solvers~\cite{Johnson1978}. Both $\theta(r)$ and the classical Pr\"ufer angle describe the orientation of the phase-space state vector $(\psi, p)$, differing only by convention-dependent angular shifts and sign choices. Consequently, the winding-number quantization introduced by Molano and Arango~\cite{Molano2021} may be interpreted as a topological counting of generalized Pr\"ufer phase rotations along the Schr\"odinger flow. The results summarized below therefore provide a rigorous Sturm--Liouville foundation for the geometric phase-space quantization procedure.

Equation~\eqref{eq:radial_schrodinger} belongs to the quasi-derivative Sturm--Liouville operator class analyzed via generalized Pr\"ufer coordinates by Bandyopadhyay, \c{C}etinkaya, and Cuchta~\cite{Bandyopadhyay2025}. The unreduced radial Schr\"odinger equation may be cast into the quasi-derivative Sturm--Liouville form with coefficient $r^2$ and shift function $s(r)=0$ in spherical coordinates. Their generalized Pr\"ufer analysis establishes three foundational guarantees for the Molano--Arango phase-space formulation~\cite{Molano2021}:
\begin{enumerate}

\item \emph{Exact Bijective Equivalence:} Theorem~2.1 of Ref.~\cite{Bandyopadhyay2025} establishes a one-to-one correspondence between nontrivial $C^2$ solutions of Eq.~\eqref{eq:radial_schrodinger} and solutions of the first-order polar phase-amplitude system with $R(r)\neq0$.

\item \emph{Nodal Regularity via Lipschitz Continuity:} Differentiating $R(r)$ and $\theta(r)$ along Eq.~\eqref{eq:phase_ode} yields separate evolution equations for the Euclidean phase-space radius and ray angle:
\begin{align}
\frac{1}{R}\frac{dR}{dr} &= \frac{1 - k^2(r; E)}{2}\sin 2\theta(r) - g(r)\sin^2\theta(r), \label{eq:polar_R_ode} \\
\frac{d\theta}{dr} &= -k^2(r; E)\cos^2\theta(r) - g(r)\sin\theta(r)\cos\theta(r) - \sin^2\theta(r). \label{eq:prufer_angle_ode}
\end{align}
By Lemma~2.2 and Corollary~2.3 of Ref.~\cite{Bandyopadhyay2025}, the right-hand side of Eq.~\eqref{eq:prufer_angle_ode} is Lipschitz continuous in $\theta$ on any regular interval $[a, b] \subset (0, \infty)$. Consequently, trajectories pass smoothly through every radial node ($\psi = 0, p \neq 0$, where $\theta = \pm\pi/2$) without the pole singularities that afflict scalar logarithmic-derivative methods.

\item \emph{Spectral Monotonicity and Winding-Number Quantization:} As proven in Theorems~3.3 and 3.6 of Ref.~\cite{Bandyopadhyay2025}, the zeros of $\psi(r)$ occur precisely at half-integer multiples of $\pi$ in $\theta(r)$ and shift strictly monotonically with respect to the spectral parameter $E$, providing the rigorous Sturm--Liouville framework underlying the winding-number interpretation of Schr\"odinger eigenstates employed in Ref.~\cite{Molano2021}.
\end{enumerate}

In classically allowed regions ($k^2(r; E) > 0$, where the local generator $\mathbf{A}(r; E)$ is elliptic), $d\theta/dr < 0$ everywhere and trajectories wind monotonically around the origin; in classically forbidden regions ($k^2(r; E) < 0$, where $\mathbf{A}(r; E)$ becomes hyperbolic), angular motion slows and is governed by the local saddle structure analyzed below. Although Eq.~\eqref{eq:phase_ode} is non-autonomous in $r$, promoting the spatial coordinate $r$ to a dynamical variable with evolution parameter $t$ ($dr/dt = 1$)---directly analogous to extended phase space in classical mechanics---lifts the flow into the three-dimensional \emph{autonomous extended phase space} $(\mathbf{x}, r) \in \mathbb{R}^2 \times \mathbb{R}$ and the two-dimensional \emph{autonomous Pr\"ufer cylinder} $(\theta, r) \in \mathbb{S}^1 \times \mathbb{R}$:
\begin{equation}
\frac{d\mathbf{x}}{dt} = \mathbf{A}(r; E)\mathbf{x}, \qquad \frac{d\theta}{dt} = F(\theta; r), \qquad \frac{dr}{dt} = 1,
\label{eq:autonomous_prufer_cylinder}
\end{equation}
where $F(\theta; r) \equiv -k^2(r; E)\cos^2\theta - g(r)\sin\theta\cos\theta - \sin^2\theta$ denotes the right-hand side of Eq.~\eqref{eq:prufer_angle_ode}. Whereas trajectories projected onto a flat two-dimensional $(\psi, p)$ plane can appear to cross as $V_{\mathrm{eff}}(r)$ varies with $r$, promoting $r$ to a state coordinate removes any explicit dependence on the evolution parameter $t$. Because the resulting two-component vector field $(F(\theta; r), 1)^\top$ on $(\theta, r)$ is autonomous in $t$, continuous in $r$, and smooth (locally Lipschitz) in $\theta$ on any regular interval $[a, b] \subset (0, \infty)$~\cite{Bandyopadhyay2025}, uniqueness of initial-value trajectories (Picard--Lindel\"of theorem) enforces two topological properties that underpin the entire propagation strategy: (i)~\emph{non-crossing and order preservation}, whereby distinct integral curves can never intersect on the $(\theta, r)$ cylinder, so any single solution curve $\theta(r)$ partitions $\mathbb{S}^1 \times \mathbb{R}$ into disjoint invariant regions and preserves strict angular ordering across any bundle of trajectories; and (ii)~\emph{time-reversal symmetry ($t \mapsto -t$)}, whereby reversing the evolution parameter inverts the vector fields to $(-\mathbf{A}(r; E)\mathbf{x}, -1)^\top$ and $(-F(\theta; r), -1)^\top$, leaving the non-intersecting integral curves geometrically unchanged while reversing their direction of traversal.

\subsection{Exact Phase-Space Propagation and Covariant Spectral Geometry}

In the autonomous extended phase space $(\mathbf{x}, r) \in \mathbb{R}^2 \times \mathbb{R}$ of Eq.~\eqref{eq:autonomous_prufer_cylinder}, each constant-$r$ plane $\mathbb{R}^2_r$ forms a local two-dimensional phase slice (or \emph{linear fiber}) governed by the instantaneous generator $\mathbf{A}(r; E)$, while transport between fibers at $r_1$ and $r_2$ is mediated by the exact state-transition matrix (linear propagator) $\mathcal{U}(r_2, r_1): \mathbb{R}^2 \to \mathbb{R}^2$~\cite{Magnus1954,Pechukas1966,Light1976,Blanes2009}, defined as the unique $2\times 2$ matrix solution to the initial-value problem
\begin{equation}
\frac{\partial}{\partial r_2}\mathcal{U}(r_2, r_1) = \mathbf{A}(r_2; E)\,\mathcal{U}(r_2, r_1), \qquad \mathcal{U}(r_1, r_1) = \mathbf{I}_2,
\label{eq:propagator_def}
\end{equation}
so that $\mathbf{x}(r_2) = \mathcal{U}(r_2, r_1)\mathbf{x}(r_1)$ for any $r_1, r_2 \in [r_{\min}, r_{\max}]$. Because $\operatorname{tr}\mathbf{A}(r; E) = -g(r)$, $\mathcal{U}(r_2, r_1)$ is an orientation-preserving linear isomorphism satisfying Liouville's determinant identity
\begin{equation}
\det\mathcal{U}(r_2, r_1) = \exp\!\left(-\int_{r_1}^{r_2} g(s)\,ds\right) > 0.
\label{eq:liouville_det}
\end{equation}

\begin{figure*}[t]
  \centering
  \includegraphics[width=0.95\textwidth]{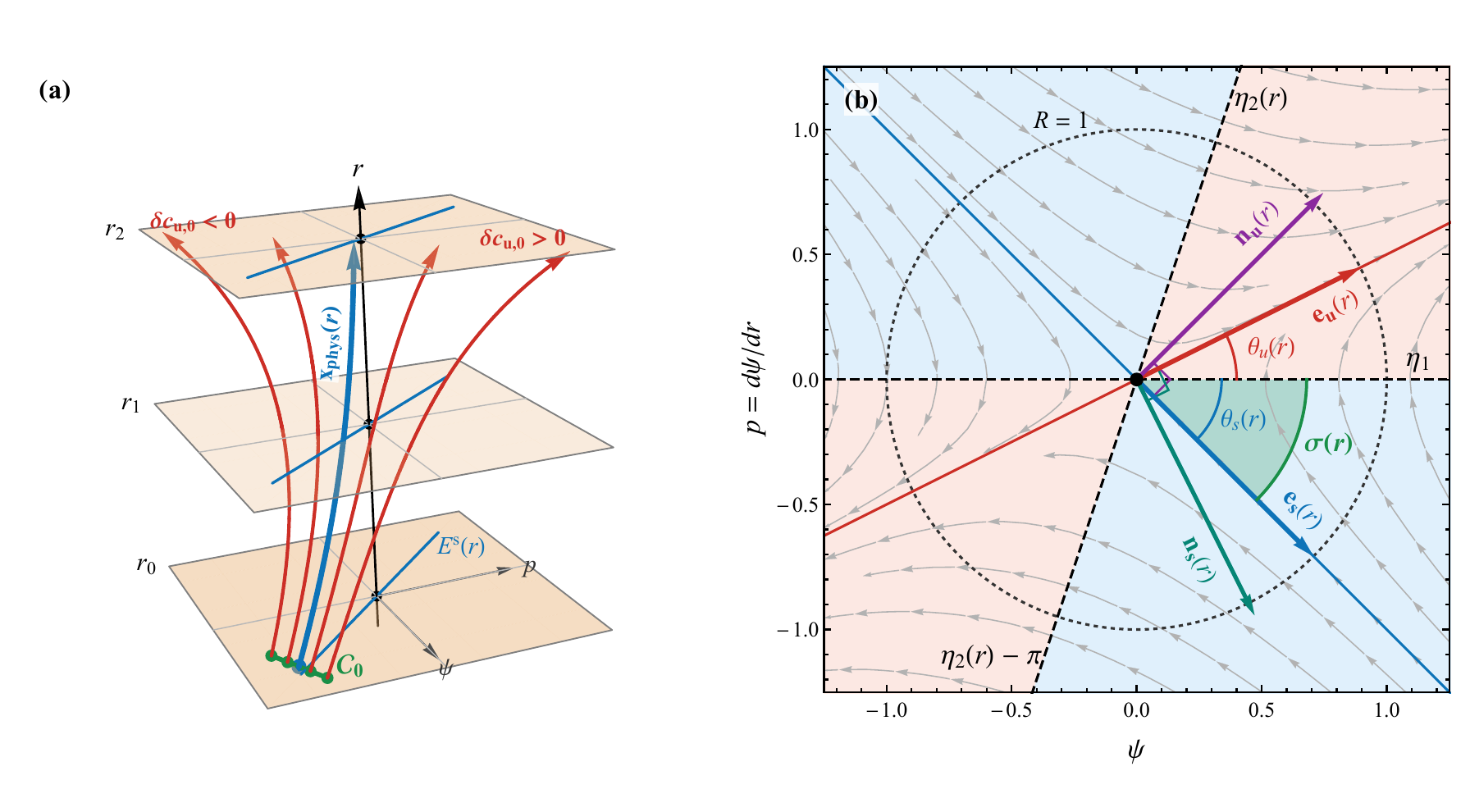}
  \caption{(a)~Three-dimensional autonomous extended phase space $(\psi, p, r) \in \mathbb{R}^2 \times \mathbb{R}$ [Eq.~\eqref{eq:autonomous_prufer_cylinder}]. Across stacked constant-$r$ fibers ($r_0 < r_1 < r_2$), the instantaneous stable subspace $E^s(r)$ (blue line) rotates continuously with $r$. By Picard--Lindel\"of uniqueness, distinct integral curves never intersect: the invariant physical decaying trajectory $(\mathbf{x}_{\mathrm{phys}}(r; E), r)$ (thick blue curve) acts as an impenetrable invariant ordering boundary inside the collapsing corridor, separating surrounding trajectories (red curves) initialized along a transverse chord $\mathcal{C}_0$ (green) with $\delta c_{u,0} > 0$ and $\delta c_{u,0} < 0$ that peel toward opposite expanding sectors in forward parameter $+t$ and funnel back onto $\mathbf{x}_{\mathrm{phys}}(r; E)$ under time reversal $t \mapsto -t$. (b)~Local phase-plane stream geometry on a constant-$r$ fiber $\mathbb{R}^2_r$ in the classically forbidden regime ($k^2(r; E) < 0$, $g(r) > 0$). The zero-radial-velocity nullclines at angles $\eta_1 \in \{0, \pi\}$ and $\eta_2(r)$ [dashed lines, Eq.~\eqref{eq:nullclines}] partition the fiber into collapsing stream sectors ($dR/dr < 0$, blue) and expanding stream sectors ($dR/dr > 0$, pink). Streamlines peel away from the angularly repelling contravariant stable eigenvector $\mathbf{e}_s(r)$ at angle $\theta_s(r) < 0$ (blue arrow), cross the angular safety margin $\sigma(r)$ [green wedge, Eq.~\eqref{eq:safety_margin}] into the expanding sector, and converge onto the attracting unstable eigenvector $\mathbf{e}_u(r)$ at angle $\theta_u(r) > 0$ (red arrow). The reciprocal covariant dual covectors $\mathbf{n}_s(r) \perp \mathbf{e}_u(r)$ (teal arrow) and $\mathbf{n}_u(r) \perp \mathbf{e}_s(r)$ (violet arrow) [Eq.~\eqref{eq:covariant_eigenvectors}] carry Euclidean norm $\|\mathbf{n}_{s,u}(r)\|_2 = \csc\Delta\theta(r) > 1$ extending past the unit rescaling circle $R = 1$ (dotted), enabling leakage-free extraction of $c_s(r)$ and $c_u(r)$ [Eq.~\eqref{eq:leakage_free_projection}].}
  \label{fig:stream_geometry}
\end{figure*}

In a classically forbidden region where $k^2(r; E) < 0$ (with local evanescent decay rate $|k(r; E)| \equiv \sqrt{-k^2(r; E)} > 0$), Eqs.~\eqref{eq:polar_R_ode}--\eqref{eq:prufer_angle_ode} define an $r$-dependent stream field in $(\psi, p)$ whose geometry deforms continuously across fibers. As illustrated in Fig.~\ref{fig:stream_geometry}(b), at any frozen coordinate $r$ the sign of $dR/dr$ in Eq.~\eqref{eq:polar_R_ode} partitions the fiber $\mathbb{R}^2_r$ into collapsing stream sectors ($dR/dr < 0$, where flow vectors point inward toward the origin) and expanding stream sectors ($dR/dr > 0$, where flow vectors point outward), separated by two zero-radial-velocity nullclines at angles
\begin{equation}
\eta_1 \in \{0, \pi\}, \qquad \cot\eta_2(r) = \frac{g(r)}{1 - k^2(r; E)},
\label{eq:nullclines}
\end{equation}
where the principal branch $\eta_2(r) \in (0, \pi/2]$ is taken for $g(r) \ge 0$, smoothly yielding $\eta_2(r) = \pi/2$ when $g(r) = 0$.

Freezing the instantaneous generator $\mathbf{A}(r; E)$ on a fiber in the classically forbidden region yields two distinct real hyperbolic eigenvalues of opposite sign,
\begin{equation}
\lambda_{s,u}(r; E) = -\frac{g(r)}{2} \mp \sqrt{\frac{g^2(r)}{4} - k^2(r; E)}, \qquad \lambda_s(r; E) < 0 < \lambda_u(r; E),
\label{eq:eigenvalues}
\end{equation}
separated by the spectral gap
\begin{equation}
\Lambda_\theta(r) \equiv \lambda_u(r; E) - \lambda_s(r; E) = 2\sqrt{\frac{g^2(r)}{4} - k^2(r; E)} > 0,
\label{eq:spectral_gap}
\end{equation}
whose reciprocal $\Lambda_\theta(r)^{-1}$ defines the characteristic local hyperbolic length scale over which the two modes separate exponentially. The corresponding unit right (contravariant) eigenvectors of $\mathbf{A}(r; E)$ are
\begin{equation}
\mathbf{e}_{s,u}(r) = \frac{1}{\sqrt{1 + \lambda_{s,u}^2(r; E)}}\begin{pmatrix} 1 \\ \lambda_{s,u}(r; E) \end{pmatrix} = \begin{pmatrix} \cos\theta_{s,u}(r) \\ \sin\theta_{s,u}(r) \end{pmatrix},
\label{eq:eigenvectors}
\end{equation}
which span the local stable (decaying) and unstable (growing) fiber subspaces $E^{s,u}(r) \equiv \mathrm{span}\{\mathbf{e}_{s,u}(r)\}$ at projective ray angles
\begin{equation}
\theta_s(r) \equiv \arctan\lambda_s(r; E) \in \left(-\frac{\pi}{2}, 0\right), \qquad \theta_u(r) \equiv \arctan\lambda_u(r; E) \in \left(0, \frac{\pi}{2}\right).
\label{eq:subspace_angles}
\end{equation}

On each constant-$r$ fiber, any propagated state vector $\mathbf{x}(r) \in \mathbb{R}^2_r$ decomposes uniquely along the instantaneous contravariant eigenbasis $\mathbb{R}^2_r = E^s(r) \oplus E^u(r)$ as
\begin{equation}
\mathbf{x}(r) = c_s(r)\mathbf{e}_s(r) + c_u(r)\mathbf{e}_u(r),
\label{eq:modal_decomposition}
\end{equation}
where $c_s(r)$ and $c_u(r)$ are the oblique coordinates measured from the origin along $E^s(r)$ and $E^u(r)$. In the slow-frame-rotation (frozen-frame) limit---where $k^2(r; E)$ and $g(r)$ vary negligibly over the local hyperbolic length scale $\Lambda_\theta(r)^{-1}$ [Eq.~\eqref{eq:spectral_gap}]---a decaying state remains aligned with $E^s(r)$ ($c_u(r) = 0$), whereas finite-rate spatial variation and discrete propagation induce a nonzero transverse component $c_u(r)$ along $E^u(r)$. Because $\mathbf{A}(r; E)$ is non-symmetric, its right eigenvectors form an oblique frame with angular separation $\Delta\theta(r) \equiv \theta_u(r) - \theta_s(r) \in (0, \pi)$ and overlap $\langle \mathbf{e}_s(r), \mathbf{e}_u(r) \rangle = \cos\Delta\theta(r) \neq 0$, so ordinary dot products couple both coordinates:
\begin{equation}
\langle \mathbf{e}_s(r), \mathbf{x}(r) \rangle = c_s(r) + c_u(r)\cos\Delta\theta(r), \qquad \langle \mathbf{e}_u(r), \mathbf{x}(r) \rangle = c_s(r)\cos\Delta\theta(r) + c_u(r).
\label{eq:naive_projection_leakage}
\end{equation}
In particular, $\langle \mathbf{e}_u(r), \mathbf{x}(r) \rangle = c_s(r)\cos\Delta\theta(r) \neq 0$ even along the instantaneous stable axis $E^s(r)$ ($c_u(r) = 0$).

Confronting $\mathbf{x}(r)$ cleanly against $E^s(r)$ therefore requires the reciprocal biorthogonal basis~\cite{MorseFeshbach1953,Horn2012}. Assembling $\mathbf{V}(r) \equiv [\mathbf{e}_s(r)\;\mathbf{e}_u(r)]$, the rows of $\mathbf{V}(r)^{-1}$ define the left (covariant) eigenvectors $\mathbf{n}_s^\top(r)$ and $\mathbf{n}_u^\top(r)$ satisfying $\mathbf{n}_\alpha^\top(r)\mathbf{A}(r; E) = \lambda_\alpha(r; E)\mathbf{n}_\alpha^\top(r)$ and $\langle \mathbf{n}_\alpha(r), \mathbf{e}_\beta(r) \rangle = \delta_{\alpha\beta}$ ($\alpha, \beta \in \{s, u\}$):
\begin{equation}
\mathbf{n}_s(r) = \csc\Delta\theta(r)\begin{pmatrix} \sin\theta_u(r) \\ -\cos\theta_u(r) \end{pmatrix}, \qquad
\mathbf{n}_u(r) = \csc\Delta\theta(r)\begin{pmatrix} -\sin\theta_s(r) \\ \cos\theta_s(r) \end{pmatrix}.
\label{eq:covariant_eigenvectors}
\end{equation}
Pairing Eq.~\eqref{eq:modal_decomposition} with this left eigenbasis recovers the exact oblique coordinates $\boldsymbol{\xi}(r) \equiv (c_s(r), c_u(r))^\top = \mathbf{V}(r)^{-1}\mathbf{x}(r)$:
\begin{equation}
c_s(r) = \langle \mathbf{n}_s(r), \mathbf{x}(r) \rangle, \qquad
c_u(r) = \langle \mathbf{n}_u(r), \mathbf{x}(r) \rangle.
\label{eq:leakage_free_projection}
\end{equation}
As depicted in Fig.~\ref{fig:stream_geometry}(b), biorthogonality requires $\|\mathbf{n}_{s,u}(r)\|_2 = \csc\Delta\theta(r)$, which diverges as $\Delta\theta(r) \to 0$ near classical turning points ($k^2 \to 0$). Substituting the polar Pr\"ufer representation $\mathbf{x}(r) = R(r)(\cos\theta(r), \sin\theta(r))^\top$ of Eq.~\eqref{eq:ray_angle} into Eq.~\eqref{eq:leakage_free_projection} yields the fundamental bridge connecting Pr\"ufer cylinder geometry to covariant modal coordinates:
\begin{equation}
\frac{c_u(r)}{R(r)} = \frac{\langle \mathbf{n}_u(r), \mathbf{x}(r) \rangle}{R(r)} = \csc\Delta\theta(r)\,\sin\bigl(\theta(r) - \theta_s(r)\bigr).
\label{eq:covariant_prufer_link}
\end{equation}
Equation~\eqref{eq:covariant_prufer_link} translates the scale-normalized departure $c_u(r)/R(r)$ from the instantaneous stable direction directly into the Pr\"ufer angular offset $\theta(r) - \theta_s(r)$, establishing the geometric quantity monitored and controlled by the bracketing algorithm.

Local angular linearization around the frozen equilibria $\theta_{s,u}(r)$ (derived in Appendix~\ref{app:prop1_proof}) shows that $E^s(r)$ acts as an angular repeller with divergence rate $+\Lambda_\theta(r) > 0$, whereas $E^u(r)$ acts as an angular attractor with convergence rate $-\Lambda_\theta(r) < 0$, even though radial vectors inside the collapsing sector point inward ($dR/dr < 0$). Because $\eta_2(r) - \pi \le -\pi/2 < \theta_s(r) < 0$ for all $g(r) \ge 0$, $\theta_s(r)$ lies strictly inside the collapsing corridor $\eta_2(r) - \pi < \theta(r) < 0$, separated from the bounding nullclines by the angular safety margin [green wedge in Fig.~\ref{fig:stream_geometry}(b)]
\begin{equation}
\sigma(r) \equiv \min\Bigl\{|\theta_s(r)|,\; \bigl|\theta_s(r) - (\eta_2(r) - \pi)\bigr|\Bigr\}.
\label{eq:safety_margin}
\end{equation}

When the spatial variation of $k^2(r; E)$ and $g(r)$ is restored across fibers, transforming Eq.~\eqref{eq:phase_ode} into the moving covariant modal coordinates $\mathbf{x}(r) = \mathbf{V}(r)\boldsymbol{\xi}(r)$ produces the rotating-frame differential system
\begin{equation}
\frac{d\boldsymbol{\xi}(r)}{dr} = \begin{pmatrix} \lambda_s(r; E) & 0 \\ 0 & \lambda_u(r; E) \end{pmatrix}\boldsymbol{\xi}(r) + \mathbf{M}_{\mathrm{rot}}(r)\boldsymbol{\xi}(r), \qquad \mathbf{M}_{\mathrm{rot}}(r) \equiv -\mathbf{V}(r)^{-1}\mathbf{V}'(r),
\label{eq:rotating_frame_ode}
\end{equation}
where the geometric connection matrix $(M_{\mathrm{rot}})_{\alpha\beta}(r) = -\langle \mathbf{n}_\alpha(r), \mathbf{e}_\beta'(r) \rangle$ measures the rate at which the eigenframe twists across fibers and satisfies the spectral-norm bound (derived explicitly in Appendix~\ref{app:prop1_proof})
\begin{equation}
\|\mathbf{M}_{\mathrm{rot}}(r)\|_2 \le \kappa_{\mathbf{V}}(r)\Bigl(|\theta_s'(r)| + |\theta_u'(r)|\Bigr), \qquad \kappa_{\mathbf{V}}(r) \equiv \sqrt{2}\,\csc\Delta\theta(r) < \infty
\label{eq:rotation_norm_bound}
\end{equation}
across every compact strictly hyperbolic domain $k^2(r; E) \le -k_{\min}^2 < 0$.

Equation~\eqref{eq:rotating_frame_ode} exposes the exact physical mechanism by which contamination arises across rotating fibers: the angular instability of the frozen equilibrium $E^s(r)$, quantified by $\Lambda_\theta(r) > 0$, amplifies deviations from $E^s(r)$, while spatial rotation of the eigenframe ($\theta_s'(r) \neq 0$) continuously injects growing-mode amplitude into $c_u(r)$ through the off-diagonal coupling $(M_{\mathrm{rot}})_{us}(r) = -\theta_s'(r)\csc\Delta\theta(r)$. Consequently, the instantaneous eigenspace $E^s(r)$ is not invariant under the flow. Instead, across any strictly hyperbolic interval $[r_{\mathrm{hyp}},r_{\mathrm{end}}]$ (selected by the physical asymptotic or matching ray condition at $r_{\mathrm{end}}$), non-autonomous exponential dichotomy theory~\cite{Coppel1978} guarantees a unique smooth rank-1 invariant contracting bundle $\mathcal{E}^{s}(r)=\mathrm{span}\{x_{\mathrm{phys}}(r;E)\}$, generated by a unique (up to scalar normalization) invariant decaying orbit $x_{\mathrm{phys}}(r;E)$ (an invariant ray in $\mathbb{R}^2$) with Pr\"ufer angle $\theta_{\mathrm{phys}}(r)$ whose amplitude remains bounded and contracts asymptotically across $[r_{\mathrm{hyp}},r_{\mathrm{end}}]$. Unlike the instantaneous eigenspace $E^s(r)$, the invariant contracting bundle $\mathcal{E}^s(r) = \mathrm{span}\{\mathbf{x}_{\mathrm{phys}}(r; E)\}$ is strictly preserved by the exact propagator,
\begin{equation}
\mathbf{x}_{\mathrm{phys}}(r_2; E) \propto \mathcal{U}(r_2, r_1)\mathbf{x}_{\mathrm{phys}}(r_1; E) \qquad \forall\, r_1, r_2 \in [r_{\mathrm{hyp}}, r_{\mathrm{end}}],
\label{eq:phys_ray_invariance}
\end{equation}
so the lifted curve $(\mathbf{x}_{\mathrm{phys}}(r; E), r)$---and its angular projection $\theta = \theta_{\mathrm{phys}}(r)$ on the autonomous Pr\"ufer cylinder $\mathbb{S}^1 \times \mathbb{R}$ [Eq.~\eqref{eq:autonomous_prufer_cylinder}]---constitutes an exact, invariant integral curve. Because of frame rotation, the invariant contracting bundle $\mathcal{E}^s(r)$ is displaced from the instantaneous stable eigendirection $E^s(r)$ by the leading-order rotation lag $\theta_{\mathrm{phys}}(r) - \theta_s(r) \approx \theta_s'(r)/\Lambda_\theta(r)$ dictated by Eq.~\eqref{eq:nonautonomous_drift}, and consequently retains a physically genuine modal coordinate $c_u(r) \neq 0$ whenever $\theta_s'(r) \neq 0$.

Furthermore, the invariant decaying orbit $\theta = \theta_{\mathrm{phys}}(r)$ remains strictly inside the collapsing stream corridor $\mathcal{S}_{\mathrm{coll}} \equiv \{(\theta, r) \mid \eta_2(r) - \pi < \theta < 0,\; r \in [r_{\mathrm{hyp}}, r_{\mathrm{end}}]\}$. Specifically, $dR/dr = 0$ along the bounding nullclines $\theta = 0$ and $\theta = \eta_2(r) - \pi$ (where Eq.~\eqref{eq:prufer_angle_ode} gives $d\theta/dr = -k^2(r; E) > 0$ and $d\theta/dr = -1 < 0$, respectively), while $dR/dr > 0$ throughout the adjacent expanding stream sectors. Any trajectory reaching those boundaries or entering the expanding sectors experiences non-negative radial growth ($dR/dr \ge 0$) and cannot satisfy the asymptotically contracting condition defining $\mathcal{E}^s(r)$ across $[r_{\mathrm{hyp}}, r_{\mathrm{end}}]$. Along the physical orbit inside $\mathcal{S}_{\mathrm{coll}}$, the radial amplitude therefore satisfies $dR_{\mathrm{phys}}/dr < 0$ pointwise across $[r_{\mathrm{hyp}}, r_{\mathrm{end}}]$. This geometric lift provides the order-preserving bracketing mechanism underlying GSMB [Fig.~\ref{fig:stream_geometry}(a)]. While trajectories projected onto a single two-dimensional $(\psi, p)$ plane can appear to cross as the fiber generator $\mathbf{A}(r; E)$ rotates with $r$, Picard--Lindel\"of uniqueness on the autonomous Pr\"ufer cylinder $(\theta, r)$ guarantees that \emph{no two integral curves can ever intersect}. Consequently, the invariant decaying orbit $\theta = \theta_{\mathrm{phys}}(r)$ acts as an impenetrable ordering boundary partitioning $\mathcal{S}_{\mathrm{coll}}$ into two disjoint, order-preserving invariant half-strips, as formalized in the following foundational result.

\begin{proposition}[Order-Preservation and Bracketing Theorem]
\label{prop:topological_bracketing}
For the autonomous Pr\"ufer flow~\eqref{eq:autonomous_prufer_cylinder} across a strictly hyperbolic forbidden interval $[r_{\mathrm{hyp}}, r_{\mathrm{end}}]$ ($k^2(r; E) \le -k_{\min}^2 < 0$, $g(r) \ge 0$), let $\theta_{\mathrm{phys}}(r)$ denote the invariant physical decaying orbit lying strictly inside the collapsing stream corridor $\mathcal{S}_{\mathrm{coll}} \equiv \{(\theta, r) \mid \eta_2(r) - \pi < \theta < 0,\; r \in [r_{\mathrm{hyp}}, r_{\mathrm{end}}]\}$ (where $dR_{\mathrm{phys}}/dr < 0$). Suppose two trajectories $\theta_-(r)$ and $\theta_+(r)$ enter $\mathcal{S}_{\mathrm{coll}}$ at $r_{\mathrm{hyp}}$ and exit $\mathcal{S}_{\mathrm{coll}}$ before $r_{\mathrm{end}}$ through opposite zero-radial-velocity nullclines: specifically, $\theta_+(r)$ crosses the upper nullcline $\theta = 0$ at some radius $r_+ \in (r_{\mathrm{hyp}}, r_{\mathrm{end}}]$, whereas $\theta_-(r)$ crosses the lower nullcline $\theta = \eta_2(r) - \pi$ at some radius $r_- \in (r_{\mathrm{hyp}}, r_{\mathrm{end}}]$. Then:
\begin{enumerate}
\item Throughout the pre-exit interval $r \in [r_{\mathrm{hyp}}, \min(r_-, r_+)]$, the invariant physical decaying orbit is strictly bracketed between them:
\begin{equation}
\theta_-(r) < \theta_{\mathrm{phys}}(r) < \theta_+(r).
\label{eq:topological_bracket}
\end{equation}
\item Any trajectory $\theta(r)$ initialized at $r_{\mathrm{hyp}}$ with $\theta(r_{\mathrm{hyp}}) > \theta_{\mathrm{phys}}(r_{\mathrm{hyp}})$ can exit $\mathcal{S}_{\mathrm{coll}}$ only upward across $\theta = 0$, whereas any trajectory initialized with $\theta(r_{\mathrm{hyp}}) < \theta_{\mathrm{phys}}(r_{\mathrm{hyp}})$ can exit $\mathcal{S}_{\mathrm{coll}}$ only downward across $\theta = \eta_2(r) - \pi$.
\end{enumerate}
\end{proposition}

The proof of Proposition~\ref{prop:topological_bracketing} is given in Appendix~\ref{app:prop1_proof}. 

\subsection{Finite-Step Numerical Uncertainty and the Exponential-Contamination Problem}

In any discrete numerical integration advancing a phase-space state $\tilde{\mathbf{x}}_n$ from fiber $r_n$ to fiber $r_{n+1} = r_n + \Delta r_n$, the instantaneous eigenframe rotates continuously over $[r_n, r_{n+1}]$. A discrete integrator of local order $q$ approximates the exact inter-fiber propagator $\mathcal{U}(r_{n+1}, r_n)$ up to a local truncation and floating-point roundoff defect $\boldsymbol{\tau}_n$ of magnitude $\|\boldsymbol{\tau}_n\|_2 = \mathcal{O}(\Delta r_n^{q+1}) + \mathcal{O}(\epsilon_{\mathrm{mach}})$:
\begin{equation}
\tilde{\mathbf{x}}_{n+1} = \mathcal{U}(r_{n+1}, r_n)\tilde{\mathbf{x}}_n + \boldsymbol{\tau}_n.
\label{eq:discrete_step_flow}
\end{equation}
Decomposing the numerical state on fiber $\mathbb{R}^2_{r_n}$ relative to the invariant contracting subbundle $\mathcal{E}^s(r_n) = \mathrm{span}\{\mathbf{x}_{\mathrm{phys}}(r_n; E)\}$ as $\tilde{\mathbf{x}}_n = \alpha_n \mathbf{x}_{\mathrm{phys}}(r_n; E) + \delta\mathbf{x}_n$, where $\alpha_n$ tracks the scalar amplitude along $\mathcal{E}^s(r_n)$ and $\delta\mathbf{x}_n$ denotes the transverse numerical uncertainty, the exact invariance~\eqref{eq:phys_ray_invariance} implies $\delta\mathbf{x}_{n+1} = \mathcal{U}(r_{n+1}, r_n)\delta\mathbf{x}_n + \boldsymbol{\tau}_n$. Measuring the transverse defect on each current fiber $\mathbb{R}^2_{r_n}$ defines the fiber-local covariant unstable uncertainty $\delta c_{u,n} \equiv \langle \mathbf{n}_u(r_n), \delta\mathbf{x}_n \rangle$; advancing to the arrival fiber $\mathbb{R}^2_{r_{n+1}}$ and pairing $\delta\mathbf{x}_{n+1}$ with $\mathbf{n}_u(r_{n+1})$ yields the step-to-step recurrence
\begin{equation}
\delta c_{u,n+1} = \bigl\langle \mathbf{n}_u(r_{n+1}),\, \mathcal{U}(r_{n+1}, r_n)\delta\mathbf{x}_n \bigr\rangle + \bigl\langle \mathbf{n}_u(r_{n+1}),\, \boldsymbol{\tau}_n \bigr\rangle.
\label{eq:covariant_error_recurrence}
\end{equation}

Equation~\eqref{eq:covariant_error_recurrence} isolates the two coupled mechanisms that govern a trajectory entering a classically forbidden barrier:
\begin{enumerate}
\item \emph{Turning-point defect seeding:} Even if a trajectory reaches the turning-point layer with zero prior error ($\delta\mathbf{x}_n = \mathbf{0}$), a finite step injects a covariant unstable component bounded by $|\langle \mathbf{n}_u(r_{n+1}), \boldsymbol{\tau}_n \rangle| \le \csc\Delta\theta(r_{n+1})\,\|\boldsymbol{\tau}_n\|_2$. Immediately past a turning point, where $|k(r; E)|$ is small, the eigenvector separation $\Delta\theta(r)$ is narrow ($\csc\Delta\theta(r) \gg 1$) while the eigenframe rotation rate $|\theta_s'(r)|$ peaks, generically seeding a nonzero initial covariant uncertainty $\delta c_{u,0}$ and angular departure $\delta\theta_0 \equiv \theta(r_{\mathrm{hyp}}) - \theta_{\mathrm{phys}}(r_{\mathrm{hyp}})$ on the entry fiber $r_{\mathrm{hyp}}$.
\item \emph{Hyperbolic exponential amplification:} Inside the strictly hyperbolic regime ($k^2(r; E) \le -k_{\min}^2 < 0$), the asymptotic rate estimate additionally invokes the adiabatic-hyperbolic scale-separation condition $\|\mathbf{M}_{\mathrm{rot}}(r)\|_2 \ll \Lambda_\theta(r)$ [Eq.~\eqref{eq:rotation_norm_bound}] across $[r_{\mathrm{hyp}}, r_{\mathrm{end}}]$, which physically governs the placement of $r_{\mathrm{hyp}}$ beyond the rapid-rotation turning-point layer. Under this regime assumption, the connection matrix $\mathbf{M}_{\mathrm{rot}}(r)$ in Eq.~\eqref{eq:rotating_frame_ode} contributes only subleading $\mathcal{O}(\|\mathbf{M}_{\mathrm{rot}}\|_2)$ corrections to the diagonal spectral rates $\lambda_{u,s}(r; E)$. Consequently, the first term of Eq.~\eqref{eq:covariant_error_recurrence} amplifies any existing covariant uncertainty as $\delta c_u(r) \sim \delta c_{u,0}\exp\!\bigl(\int_{r_{\mathrm{hyp}}}^r \lambda_u(s)\,ds\bigr)$, while the physical amplitude contracts as $R_{\mathrm{phys}}(r) \sim R_{\mathrm{phys}}(r_{\mathrm{hyp}})\exp\!\bigl(\int_{r_{\mathrm{hyp}}}^r \lambda_s(s)\,ds\bigr)$. Whereas individual modal amplitudes evolve at the separate rates $\lambda_u(r; E)$ and $\lambda_s(r; E)$, the projective Pr\"ufer angle $\theta(r)$ tracks their ratio [Eq.~\eqref{eq:covariant_prufer_link}] and therefore diverges at the difference of their exponents, $\Lambda_\theta(r) = \lambda_u(r; E) - \lambda_s(r; E)$ [Eq.~\eqref{eq:spectral_gap}]. On the autonomous Pr\"ufer cylinder $(\theta, r)$ [Eq.~\eqref{eq:autonomous_prufer_cylinder}], expanding $\partial_\theta F(\theta; r)$ around $\theta_s(r)$ (Appendix~\ref{app:prop1_proof}) confirms that to leading order in the rotation lag $\theta_{\mathrm{phys}}(r) - \theta_s(r) \approx \theta_s'(r)/\Lambda_\theta(r)$, the ray angle $\theta(r)$ peels away from the invariant physical decaying orbit $\theta_{\mathrm{phys}}(r)$ at the exponential rate $\Lambda_\theta(r)$:
\begin{equation}
\theta(r) - \theta_{\mathrm{phys}}(r) \approx \delta\theta_0 \exp\!\left(\int_{r_{\mathrm{hyp}}}^r \Lambda_\theta(s)\,ds\right).
\label{eq:angular_blowup}
\end{equation}
\end{enumerate}

Because $\theta = \theta_{\mathrm{phys}}(r)$ is an impenetrable invariant ordering boundary on the autonomous Pr\"ufer cylinder $(\theta, r)$ [Proposition~\ref{prop:topological_bracketing}], the sign of $\delta\theta_0$ permanently determines which nullcline boundary of the collapsing stream corridor $\mathcal{S}_{\mathrm{coll}}$ a perturbed trajectory approaches [Fig.~\ref{fig:stream_geometry}(a)]: trajectories with $\delta\theta_0 > 0$ ($\delta c_{u,0} > 0$) remain above $\theta_{\mathrm{phys}}(r)$ and can exit $\mathcal{S}_{\mathrm{coll}}$ only upward across $\theta = 0$ with angular clearance $\Delta\theta_+(r) \equiv -\theta_{\mathrm{phys}}(r)$, whereas trajectories with $\delta\theta_0 < 0$ ($\delta c_{u,0} < 0$) remain below $\theta_{\mathrm{phys}}(r)$ and can exit $\mathcal{S}_{\mathrm{coll}}$ only downward across $\theta = \eta_2(r) - \pi$ [Eq.~\eqref{eq:nullclines}] with angular clearance $\Delta\theta_-(r) \equiv \theta_{\mathrm{phys}}(r) - (\eta_2(r) - \pi)$. Any single trajectory carrying $|\delta\theta_0| > 0$ therefore possesses a leading-order \emph{escape horizon} $r_{\mathrm{esc}}$ determined implicitly by
\begin{equation}
\int_{r_{\mathrm{hyp}}}^{r_{\mathrm{esc}}} \Lambda_\theta(s)\,ds \approx \ln\!\left(\frac{\Delta\theta_{\pm}(r_{\mathrm{esc}})}{|\delta\theta_0|}\right) \qquad (\text{for } \delta\theta_0 > 0 \text{ or } \delta\theta_0 < 0, \text{ respectively}).
\label{eq:escape_horizon}
\end{equation}
For example, in a bound Coulomb tail ($V_{\mathrm{eff}}(r) = -Z/r + \ell(\ell+1)/(2\mu r^2)$, $E = -\kappa^2/(2\mu) < 0$), $\Lambda_\theta(r) \to 2\kappa$ asymptotically; even with double-precision roundoff ($|\delta\theta_0| \sim 10^{-15}$), Eq.~\eqref{eq:escape_horizon} predicts that the extrapolated departure reaches the corridor boundary at $\Delta r_{\mathrm{esc}} \approx \frac{1}{2\kappa}\ln(10^{15}) \approx 17.3/\kappa$ beyond $r_{\mathrm{hyp}}$ and subsequently exits through a nullcline into an expanding sector ($dR/dr > 0$). Once outside $\mathcal{S}_{\mathrm{coll}}$, the trajectory aligns with the angularly attracting unstable subspace $E^u(r)$ ($|c_u(r)/R(r)| = \mathcal{O}(1)$ in Eq.~\eqref{eq:covariant_prufer_link}), so its radial amplitude grows exponentially as $\exp\!\bigl(\int \lambda_u(s)\,ds\bigr)$ and overwhelms the decaying physical solution.

Whereas Proposition~1 establishes that the invariant decaying orbit $\theta_{\mathrm{phys}}(r)$ acts as an order-preserving boundary partitioning the collapsing corridor on $(\theta,r)$, Generational Stable-Manifold Bracketing (GSMB) operationalizes that ordering across discrete steps to overcome the escape-horizon barrier without modifying the radial ODE or introducing asymptotic matching approximations. At each fixed trial energy $E$, GSMB partitions $[r_{\mathrm{hyp}},r_{\mathrm{end}}]$ into adaptive sub-horizon segments shorter than $\Delta r_{\mathrm{esc}}$ and repeatedly re-subdividing a transverse phase-space chord straddling $\theta_{\mathrm{phys}}(r)$, suppressing unstable growth across the barrier and reconstructing the uncontaminated interior wavefunction via a single reverse-radius integration pass (detailed in Sec.~\ref{sec:gsmb_implementation}).

\subsection{Kato Cusp Core Initialization and Boundary-Free Radial Propagation}

Whereas generational bracketing stabilizes trajectories across classically forbidden barriers, formulating propagation natively as an initial-value evolution in the autonomous extended phase space $(\mathbf{x}, r) \in \mathbb{R}^2 \times \mathbb{R}$ [Eq.~\eqref{eq:autonomous_prufer_cylinder}] also eliminates artificial Dirichlet box walls at both endpoints $r_{\min}$ and $r_{\max}$. For systems governed by smooth, nonsingular potentials (such as nuclear motion on Born--Oppenheimer surfaces or Cartesian oscillators), standard regular Taylor or parity conditions specify the initial ray on the inner fiber $\mathbb{R}^2_{r_{\min}}$. For atomic and molecular electronic wavefunctions subject to attractive nuclear Coulomb singularities $V(r) = -Z/r + V_{\mathrm{ns}}(r)$ with $V_{\mathrm{ns}}(r)$ analytic at the origin, regular solutions in unreduced spherical coordinates ($g(r) = 2/r$) satisfy the Frobenius expansion
\begin{equation}
\psi(r) = A\,r^{\ell}\left[1 - \frac{Z\mu}{\ell+1}r + \mathcal{O}(r^2)\right] \qquad (r \to 0^+),
\label{eq:frobenius_core}
\end{equation}
where $A \neq 0$ is an arbitrary normalization constant. Whereas reduced radial coordinates $u(r) = r\psi(r)$ ($g(r) = 0$) vanish at the origin ($u(0) = 0$, corresponding to the vertical projective ray $\theta(0^+) \equiv \pm\pi/2 \pmod{\pi}$ for all $\ell \ge 0$), unreduced coordinates maintain a finite nonzero amplitude $\psi(0) = A$ for electronic $s$-wave states ($\ell = 0$), which keeps the limiting slope $p/\psi$ finite and directly defines a non-vertical core ray angle on the Pr\"ufer cylinder via Kato's cusp condition~\cite{Kato1957}:
\begin{equation}
\lim_{r\to 0^+}\frac{p(r)}{\psi(r)} = \lim_{r\to 0^+}\tan\theta(r) = -Z\mu.
\label{eq:kato_cusp}
\end{equation}
More generally, for arbitrary angular momentum $\ell \ge 0$, differentiating Eq.~\eqref{eq:frobenius_core} gives the logarithmic derivative on an inner entry fiber $r_{\min} \ll 1$:
\begin{equation}
\frac{p(r_{\min})}{\psi(r_{\min})} = \frac{\ell}{r_{\min}} - \frac{Z\mu}{\ell + 1} - \frac{\mu}{2\ell + 3}\left[2\bigl(E - V_{\mathrm{ns}}(0)\bigr) + \frac{Z^2\mu}{(\ell+1)^2}\right]r_{\min} + \mathcal{O}(r_{\min}^2),
\label{eq:kato_ray}
\end{equation}
which incorporates the centrifugal behavior $\ell/r_{\min}$ for $\ell > 0$, reduces directly to Kato's cusp condition~\eqref{eq:kato_cusp} when $\ell = 0$, and is independent of both the trial energy $E$ and the finite short-range screening offset $V_{\mathrm{ns}}(0)$ through $\mathcal{O}(1)$ (with $E$ and $V_{\mathrm{ns}}(0)$ entering only at $\mathcal{O}(r_{\min})$).

Truncating Eq.~\eqref{eq:kato_ray} at $\mathcal{O}(1)$ defines the energy-independent core launch angle $\theta_{\mathrm{core}}(r_{\min}) \in (-\pi/2, \pi/2)$ on the autonomous Pr\"ufer cylinder $\mathbb{S}^1 \times \mathbb{R}$:
\begin{equation}
\theta_{\mathrm{core}}(r_{\min}) \equiv \begin{cases}
-\arctan(Z\mu), & \ell = 0, \\[4pt]
\dfrac{\pi}{2} - \arctan\!\left(\dfrac{r_{\min}}{\ell - Z\mu r_{\min}/(\ell + 1)}\right), & \ell > 0,
\end{cases}
\label{eq:core_ray_def}
\end{equation}
where the complementary-angle form for $\ell > 0$ avoids large-slope division when $r_{\min} \ll 1$. Because $\tfrac{d}{dy}\bigl(\tfrac{\pi}{2} - \arctan(1/y)\bigr) = (1 + y^2)^{-1} = \mathcal{O}(r_{\min}^2/\ell^2)$ along $y = p/\psi = \ell/r_{\min} + \mathcal{O}(1)$, neglecting the $\mathcal{O}(r_{\min})$ energy-dependent remainder in Eq.~\eqref{eq:kato_ray} incurs an angular truncation error of only $\mathcal{O}(r_{\min}^3/\ell^2)$ for $\ell > 0$ (versus $\mathcal{O}(r_{\min})$ for $\ell = 0$, or $\mathcal{O}(r_{\min}^4/\ell^2)$ and $\mathcal{O}(r_{\min}^2)$, respectively, if the explicit $\mathcal{O}(r_{\min})$ term of Eq.~\eqref{eq:kato_ray} is retained). Equation~\eqref{eq:core_ray_def} converts the inner regularity condition into a uniquely specified launch point $(\theta_{\mathrm{core}}(r_{\min}), r_{\min})$ on $\mathbb{S}^1 \times \mathbb{R}$. Consequently, for each trial energy $E$, propagation across the interior region is performed as a forward initial-value evolution in the radial coordinate rather than by direct two-point boundary-value shooting. The discrete spectrum is then recovered globally by matching the propagated interior ray against the exterior bracketed physical angle $\theta_{\mathrm{phys}}(r_{\mathrm{hyp}};E)$.

Because $\theta_{\mathrm{core}}(r_{\min})$ in Eq.~\eqref{eq:core_ray_def} is independent of $E$ to the retained $\mathcal{O}(1)$ order, every propagation in a spectral energy scan originates from the identical initial fiber ray. Consequently, varying the trial energy does not modify the core initialization, require energy-dependent inward matching conditions, or alter the launch geometry on the Pr\"ufer cylinder. All energy dependence enters exclusively through the subsequent transport generated by $\mathbf{A}(r;E)$. This separation between local Coulomb regularity and global spectral transport streamlines spectral energy scans: every trial energy launches from the identical geometric point on the Pr\"ufer cylinder,
\begin{equation}
\mathbf{x}(r_{\min}) = \psi_0 \begin{pmatrix} 1 \\ \tan\theta_{\mathrm{core}}(r_{\min}) \end{pmatrix} = R_0 \begin{pmatrix} \cos\theta_{\mathrm{core}}(r_{\min}) \\ \sin\theta_{\mathrm{core}}(r_{\min}) \end{pmatrix},
\label{eq:core_ic}
\end{equation}
where $R_0 \equiv \psi_0\sec\theta_{\mathrm{core}}(r_{\min}) > 0$ (typically set to $R_0 = 1$) fixes the arbitrary normalization, and only the subsequent phase-space transport varies with $E$.

In comparison with finite-domain grid or basis discretizations that enforce Dirichlet conditions at $r_{\min}$ and $r_{\max}$, the Pr\"ufer initial-value formulation decouples local regularity from finite-box truncation at both boundaries. At the inner endpoint, imposing a Dirichlet condition directly on $\psi$ near the origin in unreduced coordinates is inconsistent with the nonzero $s$-wave amplitude $\psi(0) = A \neq 0$ and the Coulomb cusp~\eqref{eq:kato_cusp}; while standard grid solvers handle this by transforming to the reduced radial function $u(r) = r\psi(r)$ with $u(0) = 0$, the unreduced Pr\"ufer formulation incorporates the exact $\ell = 0$ cusp angle directly on $\mathbb{S}^1 \times \mathbb{R}$ without coordinate rescaling. At the outer endpoint, imposing a finite-box Dirichlet wall $\psi_{\mathrm{Dir}}(r_{\max}) = 0$ pins the terminal ray to $\theta(r_{\max}) = -\pi/2$; in the asymptotic tail $\psi(r) \sim c_s e^{-\kappa r} + c_u e^{+\kappa r}$, this boundary condition fixes $c_u/c_s = -e^{-2\kappa r_{\max}}$, thereby introducing a finite-box truncation perturbation $-e^{-\kappa r_{\max}}e^{+\kappa(r - r_{\max})}$ interpretable as a reflected growing component that shifts diffuse Rydberg eigenvalues unless $r_{\max}$ is extended well past the outer turning point, and discretizes the continuous spectrum ($E > 0$) into a box-normalized pseudostate representation whose spectral density converges to the physical continuum only as $r_{\max} \to \infty$. Initializing directly on $\theta_{\mathrm{core}}(r_{\min})$ [Eq.~\eqref{eq:core_ic}] and matching onto the invariant contracting bundle $\mathcal{E}^s(r)$ at $r_{\mathrm{hyp}}$ via generational bracketing (or integrating directly to the asymptotic scattering phase shift for $E > 0$), GSMB requires no artificial Dirichlet boundary conditions at either endpoint.

\subsection{Generational Stable-Manifold Bracketing: Implementation and Contraction Analysis}
\label{sec:gsmb_implementation}

\begin{figure}[htb]
  \centering
  \includegraphics[width=0.65\columnwidth]{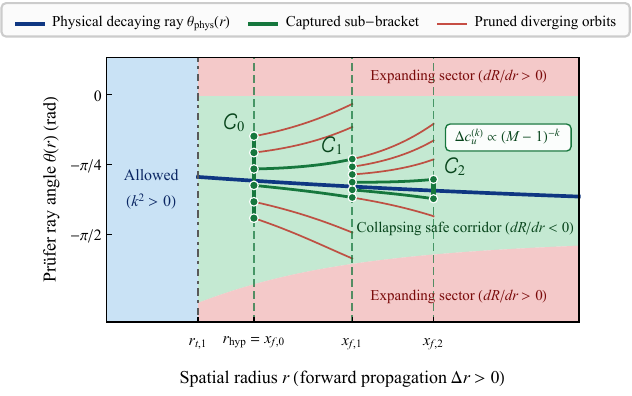}
  \caption{Forward Generational Stable-Manifold Bracketing (GSMB, $\Delta r > 0$) on the autonomous Pr\"ufer cylinder $(\theta, r)$ [Eq.~\eqref{eq:autonomous_prufer_cylinder}] for exact trajectories of Eq.~\eqref{eq:prufer_angle_ode} for a model radially confined potential ($V(r) = 0.14 r^2 - 0.45/r$, $g(r) = 2/r$) at $E = 0.388477$~a.u. To the left of the inner turning point $r_{t,1} = 2.078\,a_0$ (dot-dashed gray line), the system lies in the classically allowed region ($k^2(r; E) > 0$, sky-blue shading). Across $r_{t,1}$ in the forbidden regime ($k^2(r; E) < 0$), the zero-radial-velocity nullclines $\theta = 0$ and $\theta = \eta_2(r) - \pi$ [Eq.~\eqref{eq:nullclines}] form the upper and lower interfaces separating the safe collapsing stream corridor ($dR/dr < 0$, sage-green shading) from the upper and lower expanding sectors ($dR/dr > 0$, rose-pink shading). Inside the collapsing corridor, the physical decaying solution ray $\theta_{\mathrm{phys}}(r)$ (solid navy curve) acts as an impenetrable invariant ordering boundary between orbits peeling upward into the upper expanding sector and orbits peeling downward into the lower expanding sector. Beyond the dynamically detected hyperbolic entry radius $r_{\mathrm{hyp}} = x_{f,0}$, GSMB spawns transverse birth chords $\mathcal{C}_k$ ($M = 6$ green markers) at adaptive handoff planes $x_{f,0} < x_{f,1} < x_{f,2}$ (dashed green lines) triggered before outer trajectories (thin red curves) escape the collapsing corridor; order preservation on the Pr\"ufer cylinder maintains strict angular ordering so that at each handoff the unique adjacent sub-pair straddling $\theta_{\mathrm{phys}}(r)$ (thick green curves) is captured and re-subdivided while all outer trajectories are pruned, contracting the scale-normalized covariant unstable bracket width geometrically as $\Delta c_u^{(k)} \propto (M-1)^{-k}$.}
  \label{fig:gsmb_chords}
\end{figure}

Using the exact inter-fiber propagator $\mathcal{U}(r_2, r_1)$, covariant unstable coordinate $c_u(\mathbf{x}) = \langle \mathbf{n}_u(r), \mathbf{x} \rangle$, leading-order escape horizon $r_{\mathrm{esc}}$, and core initialization~\eqref{eq:core_ic}, Generational Stable-Manifold Bracketing (GSMB) replaces single-orbit integration across $[r_{\mathrm{hyp}}, r_{\mathrm{end}}]$ with a sequence of transverse phase-space chords $\mathcal{C}_k$ on constant-$r$ fibers that bracket $\mathcal{E}^s(r) = \mathrm{span}\{\mathbf{x}_{\mathrm{phys}}(r; E)\}$ and suppress unstable growth across the forbidden barrier (Fig.~\ref{fig:gsmb_chords}).

\noindent\textbf{Algorithm summary.}
\begin{enumerate}
\item \emph{Allowed-region sweep and hyperbolic entry detection:} Starting from Eq.~\eqref{eq:core_ic} at $r_{\min}$, integrate one trajectory across the allowed region ($k^2(r; E) > 0$) and turning point $r_{t,1}$ to detect hyperbolic entry $r_{\mathrm{hyp}} = r_{t,1} + \delta_t$ where $k^2(r; E) \le -k_{\min}^2 < 0$ ($|k(r; E)| \ge k_{\min} > 0$, keeping $\kappa_{\mathbf{V}}(r)$ bounded away from turning points). Set $r_{\mathrm{end}} = r_{\max}$ for semi-infinite bound tails, or an interior matching plane $r_{\mathrm{end}} = r_m \in (r_{t,1}, r_{t,2})$ for a barrier (integrating the exterior branch inward from $r_{\max}$ through $r_{\mathrm{hyp},2} < r_{t,2}$ to $r_m$ under $-\mathbf{A}(r; E)$).
\item \emph{Birth-chord initialization and subdivision:} At $x_{f,0} = r_{\mathrm{hyp}}$, construct straddling states $\mathbf{x}_\pm^{(0)}(x_{f,0})$ inside the collapsing corridor on opposite sides of $\theta_{\mathrm{phys}}(x_{f,0})$ (using $\theta_{\mathrm{target}}(x_{f,0}) \equiv \theta_s(x_{f,0}) + \theta_s'(x_{f,0})/\Lambda_\theta(x_{f,0})$ only as an initial predictor, with bracketing certified by opposite nullcline exits [Proposition~\ref{prop:topological_bracketing}]). At $x_{f,0}$ and each subsequent handoff fiber $x_{f,k}$, the straddling pair $\mathbf{x}_\pm^{(k)}(x_{f,k})$ spans the affine chord
\begin{equation}
  \mathcal{C}_k = \left\{ \mathbf{x}^{(k)}(w; x_{f,k}) = (1-w)\mathbf{x}_-^{(k)}(x_{f,k}) + w\mathbf{x}_+^{(k)}(x_{f,k}) \;\middle|\; w \in [0,1] \right\},
  \label{eq:birth_chord2}
\end{equation}
sampled at $M \ge 3$ uniform barycentric points $w_m = (m-1)/(M-1)$ ($m = 1,\dots,M$).
\item \emph{Adaptive handoff, sub-pair capture, and pruning:} Propagate all $M$ states forward from $x_{f,k}$ via Eq.~\eqref{eq:discrete_step_flow}. Trigger a new handoff $x_{f,k+1}$ before outer trajectories reach the nullcline safety margin $\sigma(r)$ [Eq.~\eqref{eq:safety_margin}], escape horizon $r_{\mathrm{esc}}$ [Eq.~\eqref{eq:escape_horizon}], or floor $R_{\min}$. By Pr\"ufer order preservation [Proposition~\ref{prop:topological_bracketing}], select the adjacent sub-pair $[\mathbf{x}_{m^*}^{(k+1)}, \mathbf{x}_{m^*+1}^{(k+1)}]$ straddling $\theta_{\mathrm{phys}}(x_{f,k+1})$, rescale both by $\gamma_{k+1} = \|\mathbf{x}_{m^*}^{(k+1)}\|_2^{-1}$ so $\|\mathbf{x}_-^{(k+1)}(x_{f,k+1})\|_2 = 1$, prune the remaining $M-2$ trajectories, and re-subdivide $\mathcal{C}_{k+1}$.
\item \emph{Backward stability inversion:} At $x_{f,K} = r_{\mathrm{end}}$, pick $\mathbf{x}_m^{(K)} \in \mathcal{C}_K$ and integrate backward in radius ($r: x_{f,K} \to r_{\min}$, with periodic unit-circle rescaling of $R(r)$) under $(-\mathbf{A}(r; E)\mathbf{x}, -1)^\top$ on the autonomous extended phase space [Eq.~\eqref{eq:autonomous_prufer_cylinder}, Fig.~\ref{fig:stream_geometry}(a)], reversing stability across $[r_{\mathrm{hyp}}, x_{f,K}]$ so $\theta_{\mathrm{phys}}$ attracts in reverse sweep.
\end{enumerate}

In practice, handoff radii $\{x_{f,k}\}$ are chosen adaptively at runtime via nullcline proximity, amplitude floors, and escape-horizon safeguards rather than by evaluating integral bounds or frame derivatives $(\theta_{s}^{\prime},\theta_{u}^{\prime},\kappa_{\mathbf{V}})$ directly. In Appendix~\ref{app:gsmb_proof}, Proposition~\ref{prop:gsmb_contraction} establishes rigorous geometric contraction and backward stability inversion under three formal sufficient conditions~(H1)--(H3)---covering initial straddling and sub-chord selection, uniform chord transversality and non-degeneracy, and sub-horizon handoff spacing---rather than claiming unconditional convergence for arbitrary potentials. Runtime triggers serve as practical safeguards to enforce these regimes: specifically, once $r_{\mathrm{hyp}}$ is placed past the turning-point rotation layer so that $\|\mathbf{M}_{\mathrm{rot}}(r)\|_2 \ll \Lambda_\theta(r)$ (Sec.~\ref{sec:theory}\,C), the frame-rotation integral in Eq.~\eqref{eq:handoff_spacing_hyp} is strictly subleading relative to $\int \Lambda_\theta(r)\,dr$, so triggering handoffs well inside the escape horizon $r_{\mathrm{esc}}$ [Eq.~\eqref{eq:escape_horizon}] and nullcline safety margin $\sigma(r)$ [Eq.~\eqref{eq:safety_margin}] keeps the accumulated hyperbolic exponent small enough that moderate subdivision counts (such as $M = 11$ in Sec.~\ref{sec:results}) satisfy condition~(H3) in practice.

\subsection{Two-Sided Real Wronskian Matching and Wigner Time Delay}
\label{sec:wronskian_wigner}

For bound states ($E < 0$) and open-channel scattering or tunneling resonances ($E > 0$), global quantization and phase-shift extraction are performed directly on the real energy axis without artificial box walls or complex coordinate rotation by matching interior and exterior GSMB-stabilized phase-space solutions at an interior barrier plane $r_m \in (r_{t,1}, r_{t,2})$.

In the bound-state regime ($E < 0$), outward propagation from the Kato core ray~\eqref{eq:core_ic} yields the interior state $\mathbf{x}_L(r_m; E) = (\psi_L(r_m; E), p_L(r_m; E))^\top$, while inward GSMB propagation from $r_{\max}$ yields the exponentially decaying exterior state $\mathbf{x}_R(r_m; E) = (\psi_R(r_m; E), p_R(r_m; E))^\top$. Linear dependence of the two solutions holds if and only if their phase-space Wronskian vanishes:
\begin{equation}
W[\mathbf{x}_L, \mathbf{x}_R](r_m; E) \equiv \det\bigl[\mathbf{x}_L(r_m; E)\;\mathbf{x}_R(r_m; E)\bigr] = \psi_L(r_m; E)p_R(r_m; E) - p_L(r_m; E)\psi_R(r_m; E) = 0,
\label{eq:bound_wronskian}
\end{equation}
which is equivalent to matching the projective Pr\"ufer angles $\theta_L(r_m; E) \equiv \theta_R(r_m; E) \pmod{\pi}$ while remaining completely nonsingular across radial nodes.

In the scattering regime ($E = k^2/(2\mu) > 0$) for partial wave $\ell$, the exterior asymptotic region $r \ge r_{\max}$ is spanned by two real linearly independent free-particle (Riccati--Bessel/Neumann, or regular/irregular Coulomb) asymptotic solutions $\mathbf{x}_{\mathrm{reg}}(r; E)$ and $\mathbf{x}_{\mathrm{irr}}(r; E)$ behaving as $\psi_{\mathrm{reg}}(r) \sim k^{-1}\sin(kr - \ell\pi/2 + \eta_C)$ and $\psi_{\mathrm{irr}}(r) \sim k^{-1}\cos(kr - \ell\pi/2 + \eta_C)$ (with logarithmic Coulomb phase $\eta_C = 0$ for short-range or screened potentials). Propagating both basis states inward from $r_{\max}$ to $r_m$ (using reverse-radius GSMB across the classically forbidden barrier $[r_m, r_{\mathrm{hyp},2}]$) produces two linearly independent real exterior vectors $\mathbf{x}_R^{(1)}(r_m; E)$ and $\mathbf{x}_R^{(2)}(r_m; E)$. Requiring the interior core-propagated ray $\mathbf{x}_L(r_m; E)$ to match the physical scattering combination $\cos\delta_\ell(E)\,\mathbf{x}_R^{(1)}(r_m; E) + \sin\delta_\ell(E)\,\mathbf{x}_R^{(2)}(r_m; E)$ up to an overall amplitude yields the exact real scattering phase shift
\begin{equation}
\tan\delta_\ell(E) = -\frac{\widehat{W}_{L,1}(r_m; E)}{\widehat{W}_{L,2}(r_m; E)}, \qquad
\widehat{W}_{L,j}(r_m; E) \equiv \frac{\det\bigl[\mathbf{x}_L(r_m; E)\;\mathbf{x}_R^{(j)}(r_m; E)\bigr]}{\|\mathbf{x}_L(r_m; E)\|_2\,\|\mathbf{x}_R^{(j)}(r_m; E)\|_2} \quad (j = 1, 2),
\label{eq:wronskian_tan_delta}
\end{equation}
where the scale-normalized Wronskian determinants $\widehat{W}_{L,j}$ eliminate sensitivity to exponential amplitude scaling across the barrier. At energies where $\widehat{W}_{L,2}(r_m; E_0) = 0$, the phase shift passes through $\delta_\ell(E_0) = \pi/2 \pmod{\pi}$. Differentiating the continuous unwrapped phase shift $\delta_\ell(E)$ with respect to energy yields the Smith--Wigner time delay~\cite{Wigner1955,Smith1960}, which near an isolated narrow resonance $(E_r, \Gamma)$ follows the Breit--Wigner Lorentzian profile (in atomic units, $\hbar = 1$)
\begin{equation}
\tau_W(E) \equiv 2\,\frac{d\delta_\ell(E)}{dE} = \tau_{\mathrm{bg}}(E) + \frac{\Gamma}{(E - E_r)^2 + (\Gamma/2)^2},
\label{eq:wigner_delay}
\end{equation}
where $\tau_{\mathrm{bg}}(E)$ is a smooth non-resonant background delay, $E_r$ is the resonance energy at peak time delay, and $\Gamma$ is the full width at half maximum corresponding to lifetime $\tau = 1/\Gamma$ (with $1\text{ a.u.} \approx 2.4189 \times 10^{-17}\text{ s}$).

\section{Applications and Numerical Benchmarks}
\label{sec:results}
\subsection{Coulomb Core Regularity, Escape Horizons, and Full-Domain Reconstruction in Hydrogen}
\label{sec:hydrogen_demo}

To test Coulomb core regularity, nodal smoothness, and escape-horizon control in an analytically known system, the unreduced radial hydrogen atom ($Z = 1$, $\mu = 1$, $g(r) = 2/r$) is examined starting at $r_{\min} = 10^{-4}\,a_0$. Four propagation and reference strategies are compared across identical radial intervals ($[10^{-4}, 30]\,a_0$ for $1s$ and $[10^{-4}, 45]\,a_0$ for $2s$): (i)~forward integration from a Dirichlet-like core state $\mathbf{x}(r_{\min}) = (0, 1)^\top$, (ii)~single-trajectory forward integration from the Kato cusp ray $(1, -1)^\top$, (iii)~uniform-grid finite-difference / discrete-variable-representation diagonalization with Dirichlet boundaries $u(r_{\min}) = u(r_{\max}) = 0$, and (iv)~Generational Stable-Manifold Bracketing (GSMB) followed by backward stability inversion.

\begin{figure*}[htb]
    \centering
    \includegraphics[width=\textwidth]{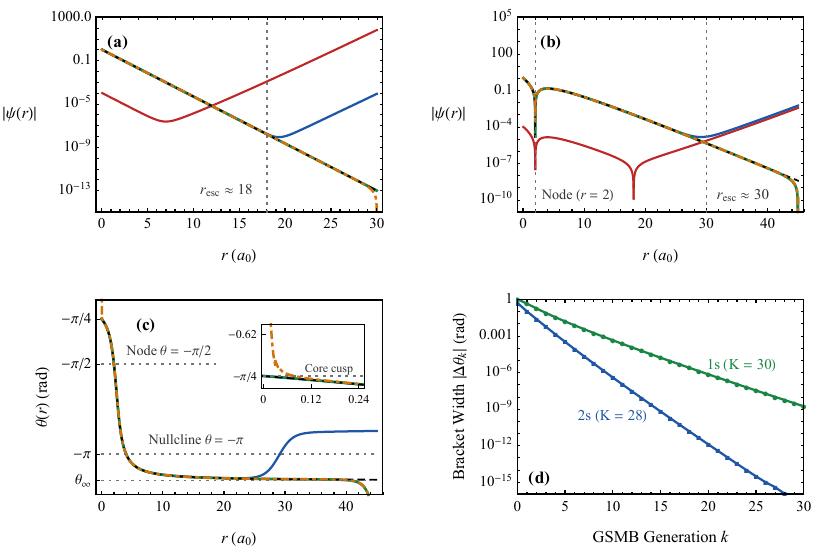}
    \caption{Benchmark comparison on unreduced radial atomic hydrogen ($Z = 1$, $\mu = 1$, $r_{\min} = 10^{-4}\,a_0$) across $[10^{-4}, 30]\,a_0$ ($1s$) and $[10^{-4}, 45]\,a_0$ ($2s$). Panels (a)--(c) compare forward integration from a Dirichlet-like core state $(0, 1)^\top$ (solid crimson), single-trajectory forward integration from the Kato cusp ray $(1, -1)^\top$ (solid navy), GSMB ($M = 11$) with backward stability inversion (solid forest green), finite-difference Dirichlet-box diagonalization enforcing $u(r_{\min}) = u(r_{\max}) = 0$ (dot-dashed amber), and the exact analytical eigenfunction (dashed black). \textbf{(a)}~$|\psi_{1s}(r)|$ at $E_{1s} = -0.5\;\mathrm{Ha}$, marking the single-trajectory escape horizon $r_{\mathrm{esc}} \approx 18\,a_0$. \textbf{(b)}~$|\psi_{2s}(r)|$ at $E_{2s} = -0.125\;\mathrm{Ha}$, showing the radial node at $r = 2\,a_0$ and $r_{\mathrm{esc}} \approx 30\,a_0$. \textbf{(c)}~Unwrapped $2s$ Pr\"ufer ray angle $\theta(r)$ and asymptotic stable angle $\theta_\infty = -\pi - \arctan(1/2)$; the inset magnifies the nuclear cusp $r \in [10^{-4}, 0.25]\,a_0$. \textbf{(d)}~Transverse bracket width $|\Delta\theta_k|$ across GSMB generations $k$ for $1s$ ($r_{\mathrm{hyp}} = 3.5\,a_0$, $K = 30$) and $2s$ ($r_{\mathrm{hyp}} = 11\,a_0$, $K = 28$).}
    \label{fig:hydrogen_gsmb}
\end{figure*}

Diagonalizing the instantaneous core generator $\mathbf{A}(r_{\min}; E)$ in Eq.~\eqref{eq:phase_ode} yields two local eigenvectors: a nearly vertical singular eigenvector $\mathbf{v}_1 \approx (-5.00025 \times 10^{-5}, 1)^\top$ aligned with the irregular $r^{-1}$ Frobenius branch, and the physical Kato cusp eigenvector $\mathbf{v}_2 = (1, -Z\mu + \mathcal{O}(r_{\min}))^\top$ [Eq.~\eqref{eq:kato_ray}]. Because $2E_{1s} + Z^2\mu = 0$ for the $1s$ ground state ($E_{1s} = -0.5\;\mathrm{Ha}$), the $\mathcal{O}(r_{\min})$ energy correction vanishes identically and $\mathbf{v}_2 = (1, -1)^\top$ is exact; shifting to the excited $2s$ state ($E_{2s} = -0.125\;\mathrm{Ha}$) alters the physical core slope by only $-3.75 \times 10^{-5}$, confirming that a single energy-independent Kato launch ray~\eqref{eq:core_ray_def} applies across an entire Rydberg series. Conversely, as shown in Fig.~\ref{fig:hydrogen_gsmb}(a)--(b) (solid crimson curves), imposing $\mathbf{x}(r_{\min}) = (0, 1)^\top$ excites the irregular $r^{-1}$ Frobenius branch, driving exponential divergence past $r \approx 7\,a_0$ for $1s$ and producing a spurious node near $r \approx 18\,a_0$ for $2s$.

Even when initialized on the exact Kato ray $\mathbf{v}_2$ at the exact eigenenergy (solid navy curves), single-trajectory forward integration encounters the hyperbolic escape horizon~\eqref{eq:escape_horizon} once finite-step truncation and roundoff seed a transverse defect $|\delta\theta_0| \sim 10^{-15}$ past the classical turning point. For $1s$ ($\kappa_{1s} = 1$, $\Lambda_\theta \to 2$), Eq.~\eqref{eq:escape_horizon} predicts $\Delta r_{\mathrm{esc}} \approx \frac{1}{2}\ln(10^{15}) \approx 17.3\,a_0$ beyond $r_{\mathrm{hyp}} = 3.5\,a_0$, and the single forward trajectory peels off the exact $e^{-r}$ tail into exponential growth around $r_{\mathrm{esc}} \approx 18\,a_0$ [Fig.~\ref{fig:hydrogen_gsmb}(a)]. For $2s$ ($\kappa_{2s} = 1/2$, $\Lambda_\theta \to 1$), the trajectory passes through the radial node at $r = 2\,a_0$ without coordinate singularities and follows $(1 - r/2)e^{-r/2}$ across five orders of magnitude before diverging near $r_{\mathrm{esc}} \approx 30\,a_0$ [Fig.~\ref{fig:hydrogen_gsmb}(b)]. By contrast, reduced-variable matrix diagonalization on the same intervals with artificial Dirichlet walls ($u(r_{\min}) = u(r_{\max}) = 0$, dot-dashed amber curves) avoids global blowup while introducing the reflected growing-mode truncation $-e^{-\kappa r_{\max}}e^{+\kappa(r - r_{\max})}$ from Sec.~\ref{sec:theory}\,D, pulling the outer tail downward within roughly $2\,a_0$ of $r_{\max} = 30\,a_0$ ($1s$) and $r_{\max} = 45\,a_0$ ($2s$).

Panel~\ref{fig:hydrogen_gsmb}(c) illustrates both inner-cusp and outer-boundary behavior geometrically on the Pr\"ufer cylinder $(\theta, r)$. Starting from the Kato cusp launch angle $\theta(r_{\min}) = -\pi/4$, the $2s$ ray angle decreases across the classically allowed region, crosses the radial node $\theta = -\pi/2$ at $r = 2\,a_0$ where Eq.~\eqref{eq:prufer_angle_ode} remains Lipschitz continuous, and enters the collapsing corridor $\mathcal{S}_{\mathrm{coll}}$ below the nullcline $\theta = -\pi$. Because the $2s$ tail decays with negative amplitude ($\psi_{2s}(r) < 0$, $\psi'_{2s}(r) > 0$ for $r > 2\,a_0$), its asymptotic stable ray lies at $\theta_\infty = \theta_s(\infty) - \pi = -\pi - \arctan(1/2) \approx -3.6052\;\mathrm{rad}$ [Eq.~\eqref{eq:subspace_angles}]. Along the single forward Kato trajectory (solid navy), angular repulsion from the invariant decaying boundary drives $\theta(r)$ upward across $\theta = -\pi$ into the expanding sector ($dR/dr > 0$) near $r \approx 28\,a_0$, approaching the attracting unstable ray $\theta_u(\infty) - \pi = -\pi + \arctan(1/2) \approx -2.6779\;\mathrm{rad}$ as the amplitude grows. Meanwhile, the reduced-variable Dirichlet-box eigenvector (dot-dashed amber) shows boundary effects at both ends: near $r_{\min}$ [Fig.~\ref{fig:hydrogen_gsmb}(c) inset], enforcing $u(r_{\min}) = 0$ mixes an irregular component $\delta\psi(r) \sim -r_{\min}/r$ whose derivative $\delta\psi'(r) \sim +r_{\min}/r^2$ dominates the finite Kato slope for $r \lesssim \sqrt{r_{\min}}$, bending $\theta_{\mathrm{Dir}}(r)$ upward from $-\pi/4$; near $r_{\max}$, the outer boundary pulls $\theta_{\mathrm{Dir}}(r)$ downward across the lower nullcline $\eta_2(r) - 2\pi$ toward $\theta(r_{\max}) = -3\pi/2 \equiv -\pi/2 \pmod{\pi}$ as $\psi_{\mathrm{Dir}}(r) \to 0^-$.

\begin{figure*}[htb]
    \centering
    \includegraphics[width=\textwidth]{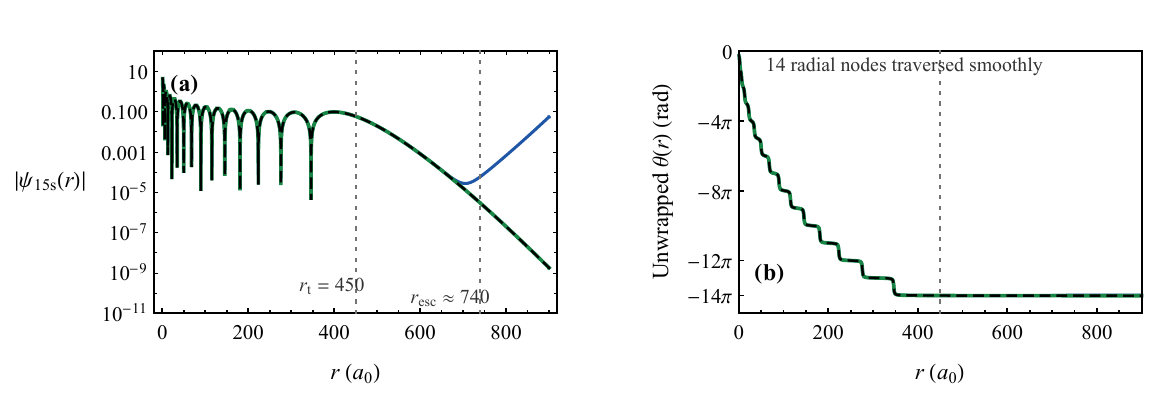}
    \caption{Multiscale nodal regularity and escape-horizon suppression for the highly excited $15s$ Rydberg state of unreduced radial atomic hydrogen ($E_{15s} = -1/450\;\mathrm{Ha}$, $r_{\min} = 10^{-4}\,a_0$, $r_{\max} = 900\,a_0$). Curves compare single-trajectory forward integration from the Kato cusp ray $(1, -1)^\top$ in double precision (solid navy), GSMB ($K = 30$, $M = 11$ across $[500, 900]\,a_0$) followed by backward stability inversion (solid forest green), and the exact analytical Laguerre--exponential eigenfunction (dashed black). \textbf{(a)}~$|\psi_{15s}(r)|$ oscillating through all $14$ radial nodes before the outer classical turning point $r_t = 450\,a_0$, showing the single-trajectory divergence past $r_{\mathrm{esc}} \approx 740\,a_0$ while GSMB remains locked onto the exact decaying tail to $r_{\max} = 900\,a_0$. \textbf{(b)}~Unwrapped Pr\"ufer ray angle $\theta(r)$ descending smoothly from $-\pi/4$ to $-14\pi - \arctan(1/15)$ across all $14$ half-integer $\pi$ node crossings without coordinate singularities.}
    \label{fig:hydrogen_rydberg}
\end{figure*}

Applying GSMB with $M = 11$ chord subdivisions across $[r_{\mathrm{hyp}}, r_{\max}]$ ($K = 30$ generations on $[3.5, 30]\,a_0$ for $1s$, and $K = 28$ generations on $[11, 45]\,a_0$ for $2s$) addresses single-orbit divergence, escape-horizon growth, and outer-wall distortion within a single real phase-space formulation. As shown in Fig.~\ref{fig:hydrogen_gsmb}(d), the transverse angular bracket width $|\Delta\theta_k|$ contracts geometrically across generations by $(M - 1)^{-1} e^{\Lambda_\theta \Delta r_k} < 1$ consistent with Proposition~\ref{prop:gsmb_contraction}, reducing angular uncertainty to $1.4 \times 10^{-9}\;\mathrm{rad}$ ($1s$) and $3.2 \times 10^{-17}\;\mathrm{rad}$ ($2s$). Subsequent backward stability inversion from $r_{\max}$ to $r_{\min}$ damps the residual outer boundary layer within roughly $2\,a_0$ of $r_{\max}$ [Fig.~\ref{fig:hydrogen_gsmb}(b)--(c)] and integrates directly in unreduced $(\psi, p)$ variables without $u/r$ division, reproducing the analytical eigenfunctions across the interior domain [Fig.~\ref{fig:hydrogen_gsmb}(c) inset] and matching the exact core slopes $p(r_{\min})/\psi(r_{\min})$ to twelve significant digits ($-1.00000000000$ for $1s$ and $-1.00003750000$ for $2s$).

To examine nodal regularity and multiscale propagation across nearly seven orders of magnitude in radius ($r \in [10^{-4}, 900]\,a_0$), the benchmark is extended to the $15s$ Rydberg state ($E_{15s} = -1/450 \approx -0.002222222222\;\mathrm{Ha}$), which has $14$ radial nodes inside its outer turning point $r_t = 450\,a_0$ [Fig.~\ref{fig:hydrogen_rydberg}]. Propagating at $E_{15s}$ from the same energy-independent Kato cusp ray $(1, -1)^\top$ at $r_{\min} = 10^{-4}\,a_0$ (with two-sided Wronskian shooting~\eqref{eq:bound_wronskian} locating $E_{15s}$ to machine precision), the unwrapped Pr\"ufer angle $\theta(r)$ decreases monotonically from $-\pi/4$ to $-14\pi - \arctan(1/15)$ across all $14$ node crossings without singularities [Fig.~\ref{fig:hydrogen_rydberg}(b)]. Beyond $r_t = 450\,a_0$, single-trajectory double-precision integration leaves the collapsing corridor near $r_{\mathrm{esc}} \approx 740\,a_0$ [Eq.~\eqref{eq:escape_horizon}], whereas GSMB ($K = 30$, $M = 11$ across $[500, 900]\,a_0$) with backward stability inversion tracks the Laguerre--exponential tail out to $r_{\max} = 900\,a_0$ [Fig.~\ref{fig:hydrogen_rydberg}(a)] and recovers the core slope $p(r_{\min})/\psi(r_{\min}) = -1.00003324$ (exact: $-1.00003319$). Because propagation starts directly on the analytical Kato cusp ray, near-origin accuracy does not depend on fine core mesh spacing, whereas grid-based finite-element or uniform finite-difference calculations remain sensitive to mesh discretization and boundary truncation across wide radial spans.

\subsection{Deep-Tunneling Real-Axis Resonance Localization, Wavefunction Reconstruction, and Local Density of States}

Whereas complex-energy approaches---such as Siegert outgoing-wave boundary conditions, complex coordinate scaling~\cite{Aguilar1971,Balslev1971,Moiseyev2011}, and complex absorbing potentials~\cite{Seideman1992,Riss1993}---represent metastable states as eigenstates of non-Hermitian operators at complex energies $E = E_r - i\Gamma/2$, shape resonances can also be examined directly on the real energy axis. Building on Molano and Arango~\cite{Molano2021}, resonance localization, orbital reconstruction, and continuum local density of states (LDOS) profiling are formulated strictly in real energy and unreduced radial phase space $\mathbf{x}(r) = (\psi(r), p(r) \equiv \psi'(r))^{\mathsf{T}} \in \mathbb{R}^2$, without complex scaling, absorbing potentials, or asymptotic Hankel and Coulomb phase-shift formulas. On the real energy axis, a shape resonance appears directly as a continuum trajectory that maximizes probability trapping inside the potential pocket relative to its exterior tail.

As shown in Fig.~\ref{fig:yukawa_gsmb_tunneling}(a) for an $\ell = 2$ screened-Coulomb centrifugal barrier ($Z_{\mathrm{eff}} = 4.35$, $r_s = 2.5\,a_0$, $\mu = 1.0$), any sub-barrier energy $0 < E < V_{\mathrm{eff}}(r_{\mathrm{peak}})$ partitions the radial coordinate across the ordered landmarks
\begin{equation}
r_{\min} < r_1(E) < r_2(E) < r_{\mathrm{peak}} < r_3(E) < r_{\mathrm{ext}} < r_{\max},
\end{equation}
where $r_{\min} = 10^{-4}\,a_0$ is the inner Kato-cusp launch radius; $r_1(E) < r_2(E) < r_3(E)$ are the three real classical turning points satisfying $V_{\mathrm{eff}}(r_i(E)) = E$, defining the classically allowed inner well $[r_1(E), r_2(E)]$ and the classically forbidden tunneling barrier $[r_2(E), r_3(E)]$ with peak at $r_{\mathrm{peak}}$; $r_{\mathrm{ext}} > r_3(E)$ marks the start of the asymptotic oscillatory region beyond the outer turning-point zone; and $r_{\max} = 240\,a_0$ spans more than four exterior de Broglie wavelengths.

To resolve narrow resonances where single-orbit forward integration across $[r_2(E), r_3(E)]$ suffers exponential contamination, the procedure follows two real-axis stages:

\begin{enumerate}
\item \textit{Stage 1: Real-axis resonance localization via minimal exterior phase-space action.} At each real trial energy $E > 0$, a trajectory starts at $r_{\min}$ along the energy-independent Kato cusp ray $\theta_{\mathrm{core}}(r_{\min})$ [\eqref{eq:core_ray_def}--\eqref{eq:core_ic}] and propagates outward across $[r_{\min}, r_{\max}]$. Normalizing each trajectory by its radial interior norm integrated up to the fixed summit $r_{\mathrm{peak}}$,
\begin{equation}
\|\psi\|_{\mathrm{well}}(E) = \left( \int_{r_{\min}}^{r_{\mathrm{peak}}} |\psi(r;E)|^2 r^2\,dr \right)^{1/2},
\end{equation}
captures both the inner well $[r_1(E), r_2(E)]$ and the inner evanescent shoulder $[r_2(E), r_{\mathrm{peak}}]$. Across the exterior oscillatory interval $[r_{\mathrm{ext}}, r_{\max}]$, the dimensionless exterior phase-space arclength is evaluated as
\begin{equation}
\mathcal{S}_{\mathrm{ext}}(E) = \frac{1}{\|\psi\|_{\mathrm{well}}(E)} \int_{r_{\mathrm{ext}}}^{r_{\max}} \sqrt{\psi(r;E)^2 + \left(\frac{p(r;E)}{k_{\mathrm{loc}}(r;E)}\right)^2}\,dr,
\end{equation}
where $k_{\mathrm{loc}}(r;E) = \sqrt{2\mu\bigl(E - V_{\mathrm{eff}}(r)\bigr)}$. Unlike the raw Euclidean phase-space length of Ref.~\cite{Molano2021}, scaling $p(r;E)$ by $k_{\mathrm{loc}}(r;E)$ makes the integrand dimensionally uniform and maps the asymptotic oscillatory ellipse $(\psi, p) \sim A\bigl(\sin(kr+\delta),\, k\cos(kr+\delta)\bigr)$ onto a circle $\sqrt{\psi^2 + (p/k_{\mathrm{loc}})^2} \approx |A|$ without $\sqrt{E}$ scaling drift. Moving off resonance (such as $E = E_r + 4\Gamma$) excites the growing barrier mode and shifts $\delta_\ell(E)$ by nearly $\pi/2$ across the Breit--Wigner transition, shifting exterior nodes and enlarging the exterior tail [red dashed curve in Fig.~\ref{fig:yukawa_gsmb_tunneling}(b)].

\begin{figure}[htbp]
  \centering
  \includegraphics[width=\textwidth]{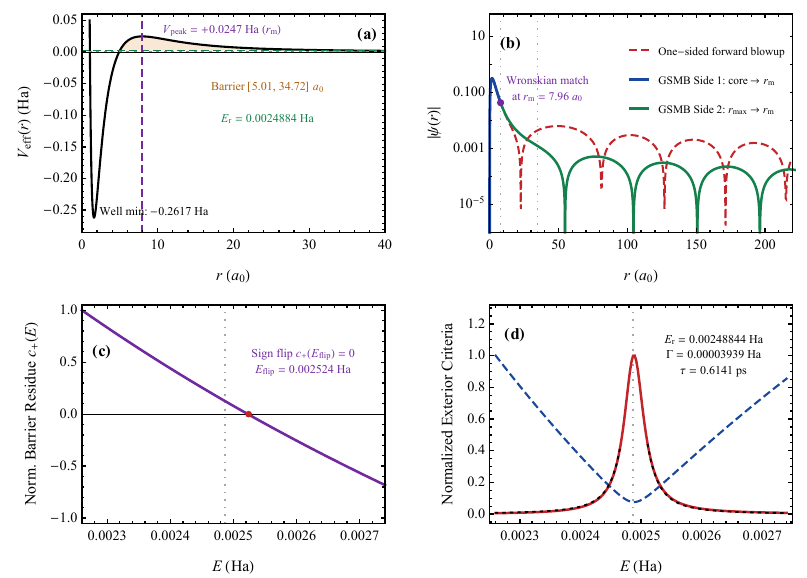}
  \caption{Real-space unreduced $(\psi, p = \psi')$ GSMB analysis of the $\ell = 2$ screened-Coulomb shape resonance ($Z_{\mathrm{eff}} = 4.35$, $r_s = 2.5\,a_0$, $\mu = 1.0$). \textbf{(a)} Effective potential $V_{\mathrm{eff}}(r)$ (solid black) showing the inner well $[r_1, r_2]$, shaded tunneling barrier $[r_2, r_3]$, resonance energy $E_r$ (green dashed), and matching plane $r_m = r_{\mathrm{peak}}$ (violet dashed). \textbf{(b)} Global orbital $|\psi(r)|$ reconstructed via two-sided unreduced GSMB matching at $r_m$ (Side~1 outward from $r_{\min}$, solid blue; Side~2 inward from $r_{\max}$, solid green), avoiding the exterior growth and off-resonant $\sim\pi/2$ phase shift of single-orbit forward integration (red dashed, $E = E_r + 4\Gamma$). \textbf{(c)} Turning-point amplitude $c_+(E) = \psi(r_3(E);E)/\|\psi\|_{\mathrm{well}}(E)$ (solid violet), crossing zero at $E_{\mathrm{flip}} = 0.002524~\mathrm{Ha}$ slightly above $E_r$. \textbf{(d)} Normalized exterior phase-space arclength $\mathcal{S}_{\mathrm{ext}}(E)$ (blue dashed), trapping power $\mathcal{S}_{\mathrm{ext}}(E)^{-2}$ (solid red), and Lorentzian fit (black dotted), yielding $E_r = 0.00248844~\mathrm{Ha}$, $\Gamma = 3.939 \times 10^{-5}~\mathrm{Ha}$, and $\tau = 0.614~\mathrm{ps}$.}
  \label{fig:yukawa_gsmb_tunneling}
\end{figure}

Figure~\ref{fig:yukawa_gsmb_tunneling}(c) plots the normalized turning-point amplitude $c_+(E) \equiv \psi(r_3(E);E)/\|\psi\|_{\mathrm{well}}(E)$ (solid violet curve), which crosses zero at $E_{\mathrm{flip}} = 0.002524~\mathrm{Ha}$ slightly above $E_r = 0.00248844~\mathrm{Ha}$. Just as a Wronskian zero $E_0$ ($\widehat{W}_{L,2}(r_m; E_0) = 0$ in Eq.~\eqref{eq:wronskian_tan_delta}) differs slightly from $E_r$, a single-point nodal condition locks the total phase shift to $\delta_\ell(E) = \delta_{\mathrm{bg}}(E) + \delta_{\mathrm{res}}(E) = \pi/2$, shifted from $\delta_{\mathrm{res}}(E_r) = \pi/2$ by the centrifugal background phase $\delta_{\mathrm{bg}}(E)$ and the turning-point offset at $r_3(E)$. Consequently, $E_{\mathrm{flip}}$ provides an initial bracket near the resonance, while minimizing the multi-wavelength exterior arclength $\mathcal{S}_{\mathrm{ext}}(E)$ [blue dashed curve in Fig.~\ref{fig:yukawa_gsmb_tunneling}(d)] reduces this background shift so that the trapping power $\mathcal{S}_{\mathrm{ext}}(E)^{-2}$ (solid red curve, black dotted Lorentzian fit) gives $E_r = 0.00248844~\mathrm{Ha}$, $\Gamma = 3.939 \times 10^{-5}~\mathrm{Ha}$, and $\tau = 1/\Gamma = 0.614~\mathrm{ps}$, in close agreement with the Siegert reference ($E_r^{\mathrm{Siegert}} = 0.00248662~\mathrm{Ha}$, $\Gamma^{\mathrm{Siegert}} = 3.708 \times 10^{-5}~\mathrm{Ha}$, $\tau^{\mathrm{Siegert}} = 0.652~\mathrm{ps}$).

\item \textit{Stage 2: Two-sided unreduced GSMB orbital reconstruction.} At a chosen resonance energy $E_r$, single-orbit forward integration from $r_{\min}$ to $r_{\max}$ amplifies $O(10^{-15})$ unstable barrier components across $[r_2(E_r), r_3(E_r)]$, distorting the exterior tail [Fig.~\ref{fig:yukawa_gsmb_tunneling}(b), red dashed curve]. Following Sec.~II\,E, propagation splits at $r_m = r_{\mathrm{peak}}$ with stepwise Euclidean renormalization ($R(r) \in [1/\gamma, \gamma]$, $\gamma = 8$): Side~1 integrates the Kato core state outward ($r_{\min} \to r_m$) to $(\hat{\psi}_L, \hat{p}_L)$ [solid blue], while Side~2 integrates the asymptotic free-wave ellipse at $r_{\max}$---$\psi_R(r_{\max};\alpha) = \cos(kr_{\max}-\alpha)$, $p_R(r_{\max};\alpha) = -k\sin(kr_{\max}-\alpha) - r_{\max}^{-1}\cos(kr_{\max}-\alpha)$ for $\alpha \in [0,\pi)$---inward from $r_{\max}$ to $r_m$, reversing the local hyperbolic instability inside $[r_m, r_3(E_r)]$. Enforcing $\hat{W}(\psi_L, \psi_R(\alpha))(r_m) = 0$ determines the matching phase $\alpha^*$ [solid green curve in Fig.~\ref{fig:yukawa_gsmb_tunneling}(b)].
\end{enumerate}

\begin{figure}[htbp]
  \centering
  \includegraphics[width=\textwidth]{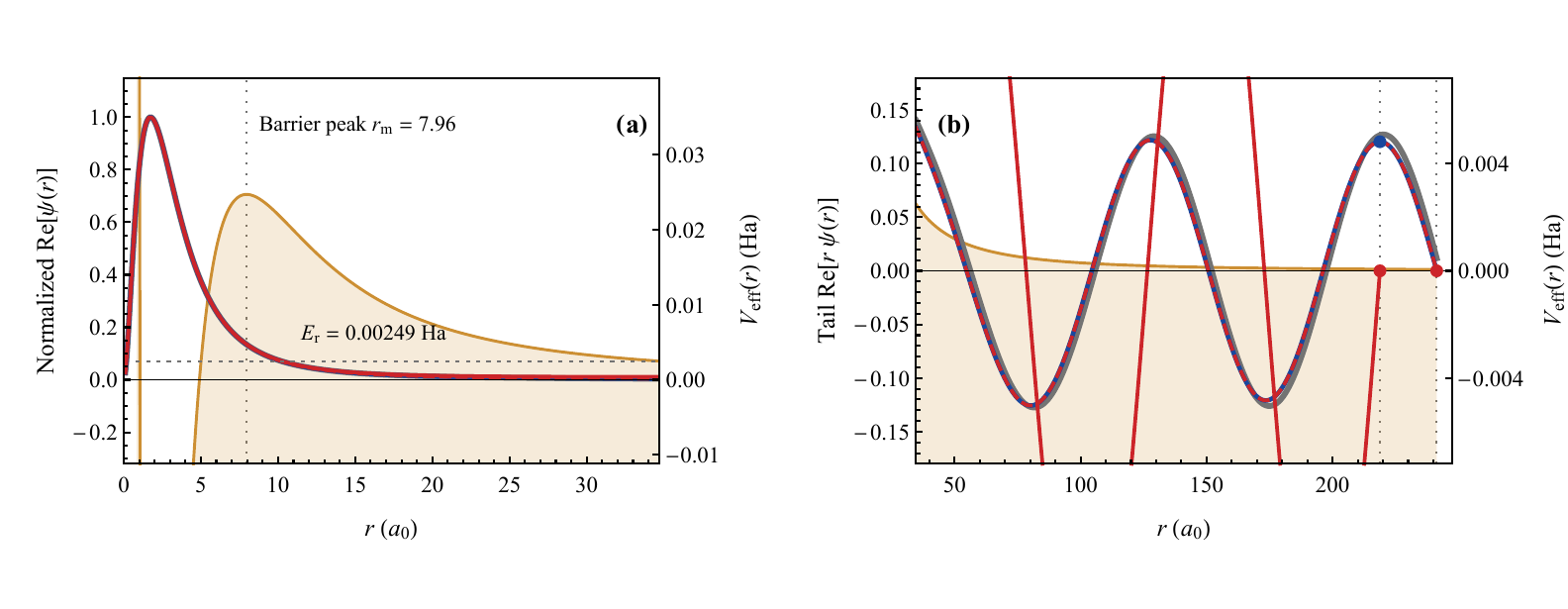}
  \caption{Spatial comparison of the reconstructed $\ell = 2$ resonant wavefunction normalized to its inner-well peak against complex-energy Siegert and finite-element Dirichlet box benchmarks, overlaid with the bottom-filled effective potential $V_{\mathrm{eff}}(r)$ (amber silhouette, right axis in Hartrees). Curves show the complex-energy outgoing Siegert reference (solid gray), the real-axis GSMB orbital (solid blue), and Hermitian \texttt{NDEigensystem} eigenstates subject to $u(r_{\mathrm{box}}) = 0$ without asymptotic endpoint guidance ($r_{\mathrm{anti}} = 218.8\,a_0$, solid red) versus with asymptotic nodal guidance from GSMB ($r_{\mathrm{node}} = 241.3\,a_0$, dashed red). \textbf{(a)} Interior and tunneling region ($0 \le r \le r_3$): all methods agree across the inner well ($r \lesssim r_m$), while unguided antinode box placement detunes $E_{\mathrm{box}}$ away from $E_r$ and excites growing-mode contamination ($c_+ \neq 0$) across the outer barrier ($r_m < r < r_3$, solid red). \textbf{(b)} Reduced exterior tail $\mathrm{Re}[r\,\psi(r)]$ ($r_3 \le r \le r_{\mathrm{node}}$), scaled by $r$ to remove $1/r$ spherical geometric dilution: guided by the GSMB radial node, \texttt{NDEigensystem} (dashed red) recovers the exact same bounded real-energy standing wave as GSMB (solid blue) across both the antinode crest (blue dot) and radial node (red dot at $r_{\mathrm{node}}$), whereas the unguided antinode box cut forces an artificial node at $u(r_{\mathrm{anti}}) = 0$ (red dot at $r_{\mathrm{anti}}$) with an inflated continuum amplitude, and the complex-energy Siegert reference (solid gray) exhibits slight asymptotic envelope growth due to $\mathrm{Im}[E] = -\Gamma/2$.}
  \label{fig:yukawa_wf_benchmark}
\end{figure}

To examine global spatial structure without assuming a priori asymptotic scattering formulas, Fig.~\ref{fig:yukawa_wf_benchmark} compares the real-axis GSMB orbital (solid blue) against the Siegert reference (solid gray) and Hermitian finite-element box eigenstates (\texttt{NDEigensystem}) imposing $u(r_{\mathrm{box}}) = r_{\mathrm{box}}\psi(r_{\mathrm{box}}) = 0$. Because an open-channel orbital oscillates asymptotically as $\psi(r) \sim A\,\sin(kr + \delta_\ell)/r$, a finite box matches a real-energy resonance most naturally when its boundary lies near an exterior node of that standing wave. Near $r_{\mathrm{box}} \sim 220\,a_0$, the box spacing $\Delta E_{\mathrm{box}} \approx \pi k / r_{\mathrm{box}} \sim 10^{-3}~\mathrm{Ha}$ is larger than $\Gamma$, so an arbitrary box placement near an antinode ($r_{\mathrm{anti}} = 218.8\,a_0$, solid red) shifts $E_{\mathrm{box}}$ away from $E_r$. While inner-well profiles agree ($r \lesssim r_m$, Fig.~\ref{fig:yukawa_wf_benchmark}(a)), this energy offset excites a growing barrier mode ($c_+ \neq 0$) across $(r_m, r_3)$ and increases the reduced exterior amplitude $\mathrm{Re}[r\,\psi(r)]$ forced to zero at $r_{\mathrm{anti}}$ [Fig.~\ref{fig:yukawa_wf_benchmark}(b), solid red]. Placing the box boundary at an exterior node indicated by GSMB ($r_{\mathrm{node}} = 241.3\,a_0$, dashed red) aligns $E_{\mathrm{box}}$ with $E_r$ and reproduces the bounded GSMB standing wave across both barrier and tail (whereas the Siegert state at $E = E_r - i\Gamma/2$ carries an outgoing imaginary part and slight envelope growth at large $r$). Two-sided GSMB incorporates this asymptotic matching directly into inward propagation from $r_{\max}$ at a single real energy $E_r$.

Beyond isolating a single pole, because $\mathcal{S}_{\mathrm{ext}}(E) \approx (r_{\max} - r_{\mathrm{ext}})\,|A(E)|/\|\psi\|_{\mathrm{well}}(E)$ measures the asymptotic standing-wave amplitude per unit interior norm, the Stage~1 trajectory directly yields the energy-normalized 1D partial interior LDOS integrated over the potential pocket,
\begin{equation}
\rho_{\mathrm{well}}^{(\ell)}(E) = \int_{r_{\min}}^{r_{\mathrm{peak}}} \bigl|u_E^{(\ell)}(r)\bigr|^2\,dr = \frac{2\mu}{\pi k}\,\frac{\|\psi\|_{\mathrm{well}}^2(E)}{|A(E)|^2} \simeq \frac{2\mu (r_{\max} - r_{\mathrm{ext}})^2}{\pi k\,\mathcal{S}_{\mathrm{ext}}(E)^2},
\label{eq:ldos_well_gsmb}
\end{equation}
where $u_E^{(\ell)}(r) = r\psi_E^{(\ell)}(r)$ is normalized on the energy scale ($\int_0^\infty u_E^{(\ell)}(r)u_{E'}^{(\ell)}(r)\,dr = \delta(E - E')$), and the total 3D interior LDOS is $\rho_{\mathrm{well}}(E) = \sum_{\ell=0}^{\ell_{\max}} (2\ell + 1)\,\rho_{\mathrm{well}}^{(\ell)}(E)$. Near a narrow sub-barrier pole, $\rho_{\mathrm{well}}^{(\ell)}(E)$ shares the Breit--Wigner Lorentzian profile of $\Delta\rho_\ell(E) = \pi^{-1}\,d\delta_\ell/dE = \tau(E)/(2\pi)$, while remaining well defined above the barrier peak $V_{\mathrm{peak}}^{(\ell)} \equiv V_{\mathrm{eff}}(r_{\mathrm{peak}}^{(\ell)})$ when turning points disappear.

\begin{figure}[htbp]
  \centering
  \includegraphics[width=\textwidth]{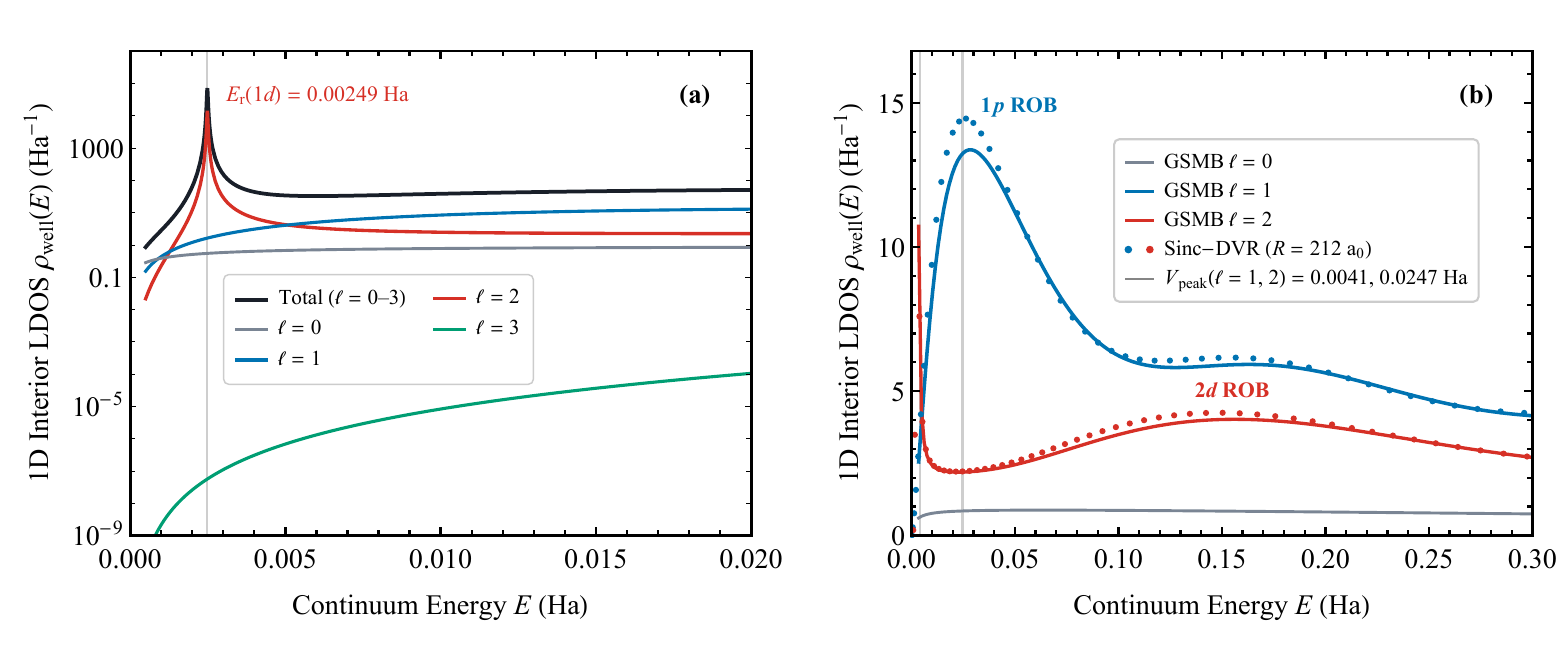}
 \caption{Continuous real-axis GSMB interior $\rho_{\mathrm{well}}^{(\ell)}(E)$ [Eq.~\eqref{eq:ldos_well_gsmb}] for the screened-Coulomb potential ($Z_{\mathrm{eff}} = 4.35$, $r_s = 2.5\,a_0$, $\mu = 1.0$). \textbf{(a)} Sub-barrier and low-energy regime ($E \in [0, 0.020]\;\mathrm{Ha}$) on a logarithmic scale, showing the total 3D interior LDOS summed over $\ell = 0\text{--}3$ (solid black) and its partial-wave decomposition into $\ell = 2$ (solid red, dominated by the narrow $1d$ shape resonance at $E_r = 0.00249\;\mathrm{Ha}$), $\ell = 1$ (solid blue), $\ell = 0$ (solid gray), and $\ell = 3$ (solid green). \textbf{(b)} Over-the-barrier regime ($E \in [0, 0.30]\;\mathrm{Ha}$) on a linear scale comparing continuous GSMB curves for $\ell = 1$ (solid blue), $\ell = 2$ (solid red), and $\ell = 0$ (solid gray) against discrete box-discretized Colbert--Miller Sinc-DVR eigenstates ($R_{\mathrm{box}} = 212\,a_0$, filled circles). Vertical gray lines mark the centrifugal barrier summits $V_{\mathrm{peak}}^{(\ell=1)} = 0.0041\;\mathrm{Ha}$ and $V_{\mathrm{peak}}^{(\ell=2)} = 0.0247\;\mathrm{Ha}$, highlighting the broad $1p$ and $2d$ over-the-barrier probability arches.}
  \label{fig:yukawa_ldos_rob}
\end{figure}

Figure~\ref{fig:yukawa_ldos_rob} displays sub-barrier tunneling and over-the-barrier (ROB) continuum structures on a common real-energy basis. For $E \in (0, 0.020]\;\mathrm{Ha}$ [Fig.~\ref{fig:yukawa_ldos_rob}(a)], the total interior LDOS ($\ell = 0\text{--}3$, solid black) is dominated by the narrow $\ell = 2$ peak at $E_r(1d) = 0.00249\;\mathrm{Ha}$ (solid red), nearly five orders of magnitude above the $\ell = 0, 1$ backgrounds while $\ell = 3$ is suppressed by its higher barrier. Above $V_{\mathrm{peak}}^{(\ell=1)} = 0.0041\;\mathrm{Ha}$ and $V_{\mathrm{peak}}^{(\ell=2)} = 0.0247\;\mathrm{Ha}$ [$E \in (0, 0.30]\;\mathrm{Ha}$, Fig.~\ref{fig:yukawa_ldos_rob}(b)], broad $1p$ ($\ell = 1$, peaking near $E \approx 0.028\;\mathrm{Ha}$) and $2d$ ($\ell = 2$, centered near $E \approx 0.15\;\mathrm{Ha}$) ROB features appear, separated from the $1d$ resonance tail by a minimum near $V_{\mathrm{peak}}^{(\ell=2)}$.

Discrete interior weights $w_m / \Delta E_m$ from 1D Colbert--Miller Sinc-DVR diagonalizations~\cite{Colbert1992} on $(0, R_{\mathrm{box}}]$ ($R_{\mathrm{box}} = 212\,a_0$, $N = 870$, $w_m = \int_0^{r_{\mathrm{peak}}^{(\ell)}} |u_m^{(\ell)}(r)|^2\,dr$, $\Delta E_m = (E_{m+1} - E_{m-1})/2$) track the continuous GSMB curves across both ROB features [Fig.~\ref{fig:yukawa_ldos_rob}(b)]. At the same time, fixed-box discrete energies $E_m$ depend on $R_{\mathrm{box}}$, and at low energy the box spacing $\Delta E_{\mathrm{box}} \propto \sqrt{E}/R_{\mathrm{box}} \sim 5 \times 10^{-4}\;\mathrm{Ha}$ is wider than $\Gamma \approx 3.9 \times 10^{-5}\;\mathrm{Ha}$, making continuous real-axis evaluation convenient for resolving narrow peaks without sweeping box sizes.

This purely real two-stage formulation provides direct physical visibility on the real energy axis, exact Kato core launching without artificial Dirichlet walls, no asymptotic special-function formulas, and control over barrier growth. In practice, low-energy resonances benefit from an exterior interval spanning several de Broglie wavelengths ($r_{\max} \gtrsim 3\lambda\text{--}4\lambda$ past $r_3$) to resolve the $\mathcal{S}_{\mathrm{ext}}(E)$ minimum, while narrow states ($\Gamma \ll E_r$) benefit from refined energy sampling (with forward GSMB rescaling in Stage~1 when barrier attenuation exceeds double precision).

\section{Conclusions}
\label{sec:conclusions}

Generational Stable-Manifold Bracketing (GSMB) has been formulated and examined as a purely real, initial-value phase-space propagator for one-dimensional, single-channel bound states and tunneling shape resonances without complex coordinate rotation, complex absorbing potentials, or artificial Dirichlet box boundaries. By combining generalized Pr\"ufer equivalence, energy-independent Kato cusp initialization, order-preserving trajectory bracketing on the autonomous Pr\"ufer cylinder with retroactive backward stability inversion, and two-sided real Wronskian matching with Wigner time delay, the formulation addresses exponential barrier instabilities directly on the real energy axis. Crucially, identifying the physical decaying solution as an impenetrable invariant ordering boundary transforms an exponentially ill-conditioned asymptotic shooting problem into a local, topologically certified enclosure in unreduced phase space---preserving linear superposition, exact Wronskians, and nodal regularity without transforming to nonlinear Riccati ratios or truncating the physical domain. In practice, handoff radii rely on nullcline margins, amplitude floors, and escape-horizon estimates, whereas Appendix~\ref{app:gsmb_proof} establishes the analytical sufficient condition~(H3) under which frame-rotation contraction and backward stability inversion are guaranteed.

Across unreduced atomic hydrogen spanning $1s$ and $2s$ states as well as the $15s$ Rydberg state across nearly seven orders of magnitude in radius and an $\ell = 2$ screened-Coulomb shape resonance, these benchmarks test invariant ordering, escape-horizon control, and backward reconstruction while launching directly from the exact Kato cusp ray without near-origin mesh dependence, preserving nodal regularity, and recovering twelve-digit nuclear cusp slopes. On the real energy axis, exterior phase-space arclength localization, two-sided Wronskian matching, and interior local density of states yield resonance positions, widths, lifetimes, and over-the-barrier features without non-Hermitian continuation or arbitrary box-wall tuning.

This framework establishes the one-dimensional, single-channel foundation of GSMB from which coupled-channel and multidimensional extensions can be developed. Compared with Riccati and log-derivative propagators that encounter singularities at wavefunction nodes ($\det\boldsymbol{\Psi} = 0$) and typically require diabatic formulations, unreduced phase-space propagation crosses nodes smoothly and accommodates first-derivative couplings directly. Extending GSMB from a single invariant boundary in $\mathbb{R}^2$ to coupled multi-channel and two-degree-of-freedom systems via higher-dimensional frames and subspace stabilization on Stiefel or Grassmannian manifolds provides a natural direction for future work.

\section*{Acknowledgments}
Financial support was provided through internal research grants from Universidad Icesi. 

\section*{Conflicts of Interest}
The author declares no conflicts of interest.

\section*{AI Use Disclosure}
The underlying geometric insight, physical formulation, and Generational Stable-Manifold Bracketing (GSMB) method were conceived by the author. During the preparation of this manuscript, the author used Google Gemini to assist in casting these geometric concepts into conventional mathematical form, structuring and verifying formal proofs, refactoring numerical benchmark scripts, and formatting \LaTeX\ equations. All derivations, proofs, and numerical results were directed, reviewed, and independently verified by the author against exact analytical solutions and reference benchmarks, and the author assumes full responsibility for the final manuscript.

\section*{Data and Code Availability}
The Mathematica scripts and numerical benchmark data supporting this study---including the GSMB propagation and two-sided matching routines, finite-element and sinc-DVR reference scripts, radial hydrogen and screened-Coulomb resonance benchmarks, and zero-energy crossover calculations---are available from the author upon reasonable request.

\appendix

\section{Covariant Frame Algebra, Angular Stability, and Proof of Proposition~\ref{prop:topological_bracketing}}
\label{app:prop1_proof}

Expanding the angular right-hand side $F(\theta; r) = -k^2(r; E)\cos^2\theta - g(r)\sin\theta\cos\theta - \sin^2\theta$ of Eq.~\eqref{eq:prufer_angle_ode} to first order in $\delta\theta \equiv \theta - \theta_{s,u}(r)$ around the frozen fiber equilibria $F(\theta_{s,u}(r); r) = 0$ gives
\begin{equation}
\left.\frac{\partial F(\theta; r)}{\partial\theta}\right|_{\theta_{s,u}(r)} = -\bigl[2\lambda_{s,u}(r; E) + g(r)\bigr] = \pm\Lambda_\theta(r),
\label{eq:angular_stability}
\end{equation}
establishing that $E^s(r)$ is an angular repeller ($+\Lambda_\theta(r) > 0$) while $E^u(r)$ is an angular attractor ($-\Lambda_\theta(r) < 0$). When spatial variation is restored across fibers, differentiating $\delta\theta(r) \equiv \theta(r) - \theta_s(r)$ along Eq.~\eqref{eq:prufer_angle_ode} yields the inhomogeneous drift equation
\begin{equation}
\frac{d(\delta\theta)}{dr} = \Lambda_\theta(r)\,\delta\theta(r) - \frac{d\theta_s(r)}{dr} + \mathcal{O}(\delta\theta^2), \qquad \frac{d\theta_s(r)}{dr} = \frac{1}{1 + \lambda_s^2(r; E)}\frac{d\lambda_s(r; E)}{dr}.
\label{eq:nonautonomous_drift}
\end{equation}
Furthermore, using the trigonometric identity $\partial_\theta^2 F(\theta; r) = -4F(\theta; r) - 2\bigl[1 + k^2(r; E)\bigr]$ at $F(\theta_s(r); r) = 0$ together with Eq.~\eqref{eq:angular_stability}, Taylor expansion of $\partial_\theta F(\theta; r)$ around $\theta_s(r)$ yields
\begin{equation}
\left.\frac{\partial F(\theta; r)}{\partial\theta}\right|_{\theta_{\mathrm{phys}}(r)} = \Lambda_\theta(r) - 2\bigl[1 + k^2(r; E)\bigr]\bigl(\theta_{\mathrm{phys}}(r) - \theta_s(r)\bigr) + \mathcal{O}\bigl((\theta_{\mathrm{phys}} - \theta_s)^2\bigr),
\label{eq:angular_taylor_phys}
\end{equation}
which governs the leading-order exponential divergence rate $\Lambda_\theta(r)$ in Eq.~\eqref{eq:angular_blowup}.

In moving covariant modal coordinates $\mathbf{x}(r) = \mathbf{V}(r)\boldsymbol{\xi}(r)$ [Eq.~\eqref{eq:rotating_frame_ode}], differentiating the unit contravariant eigenvectors $\mathbf{e}_{s,u}(r) = (\cos\theta_{s,u}(r), \sin\theta_{s,u}(r))^\top$ in Eq.~\eqref{eq:eigenvectors} and pairing with the biorthogonal left covectors $\mathbf{n}_{s,u}(r)$ in Eq.~\eqref{eq:covariant_eigenvectors} yields the explicit geometric connection matrix
\begin{equation}
\mathbf{M}_{\mathrm{rot}}(r) \equiv -\mathbf{V}(r)^{-1}\mathbf{V}'(r) = \csc\Delta\theta(r) \begin{pmatrix} \theta_s'(r)\cos\Delta\theta(r) & \theta_u'(r) \\ -\theta_s'(r) & -\theta_u'(r)\cos\Delta\theta(r) \end{pmatrix}.
\label{eq:mrot_explicit}
\end{equation}
Direct evaluation of the Frobenius norm $\|\mathbf{M}_{\mathrm{rot}}(r)\|_F \equiv \bigl(\sum_{\alpha,\beta} |(M_{\mathrm{rot}})_{\alpha\beta}(r)|^2\bigr)^{1/2}$ from Eq.~\eqref{eq:mrot_explicit} gives
\begin{equation}
\|\mathbf{M}_{\mathrm{rot}}(r)\|_F = \csc\Delta\theta(r)\sqrt{1 + \cos^2\Delta\theta(r)}\,\sqrt{|\theta_s'(r)|^2 + |\theta_u'(r)|^2}.
\end{equation}
Applying $\|\mathbf{M}_{\mathrm{rot}}(r)\|_2 \le \|\mathbf{M}_{\mathrm{rot}}(r)\|_F$ together with $\sqrt{1 + \cos^2\Delta\theta(r)} \le \sqrt{2}$ and $\sqrt{a^2 + b^2} \le |a| + |b|$ establishes Eq.~\eqref{eq:rotation_norm_bound}.

\noindent\emph{Geometric roadmap.} On the autonomous Pr\"ufer cylinder $\mathbb{S}^1 \times \mathbb{R}$, the collapsing stream corridor $\mathcal{S}_{\mathrm{coll}}$ acts as a strip bounded by two zero-radial-velocity nullcline guardrails ($\theta = 0$ above and $\theta = \eta_2(r) - \pi$ below), while the invariant physical orbit $\theta_{\mathrm{phys}}(r)$ forms an impenetrable dividing curve strictly inside the strip. Because distinct integral curves cannot intersect on $\mathbb{S}^1 \times \mathbb{R}$, any trajectory exiting across the upper guardrail must have remained strictly above $\theta_{\mathrm{phys}}(r)$ prior to exit, and any trajectory exiting across the lower guardrail must have remained strictly below it.

\begin{proof}[Proof of Proposition~\ref{prop:topological_bracketing}]
Because $(F(\theta; r), 1)^\top$ in Eq.~\eqref{eq:autonomous_prufer_cylinder} is autonomous in $t$, continuous in $r$, and locally Lipschitz in $\theta$ on $[r_{\mathrm{hyp}}, r_{\mathrm{end}}] \subset (0, \infty)$~\cite{Bandyopadhyay2025}, Picard--Lindel\"of uniqueness guarantees that distinct integral curves never intersect on the cylinder $\mathbb{S}^1 \times \mathbb{R}$. As established in Sec.~\ref{sec:theory}\,B, the invariant decaying orbit satisfies $\eta_2(r) - \pi < \theta_{\mathrm{phys}}(r) < 0$ across $[r_{\mathrm{hyp}}, r_{\mathrm{end}}]$ (with $dR_{\mathrm{phys}}/dr < 0$ pointwise inside $\mathcal{S}_{\mathrm{coll}}$). If $\theta_+(r_{\mathrm{hyp}}) \le \theta_{\mathrm{phys}}(r_{\mathrm{hyp}})$, non-intersection forces $\theta_+(r) \le \theta_{\mathrm{phys}}(r)$ across $[r_{\mathrm{hyp}}, r_{\mathrm{end}}]$; since $\theta_{\mathrm{phys}}(r) < 0$ for all $r$, crossing the upper nullcline $\theta = 0$ is impossible, contradicting $\theta_+(r_+) = 0$ at $r_+ \in (r_{\mathrm{hyp}}, r_{\mathrm{end}}]$. Thus $\theta_+(r_{\mathrm{hyp}}) > \theta_{\mathrm{phys}}(r_{\mathrm{hyp}})$, and non-intersection preserves $\theta_+(r) > \theta_{\mathrm{phys}}(r) > \eta_2(r) - \pi$ across $[r_{\mathrm{hyp}}, r_+]$, so $\theta_+(r)$ can exit $\mathcal{S}_{\mathrm{coll}}$ only upward across $\theta = 0$. Symmetrically, if $\theta_-(r_{\mathrm{hyp}}) \ge \theta_{\mathrm{phys}}(r_{\mathrm{hyp}})$, non-intersection forces $\theta_-(r) \ge \theta_{\mathrm{phys}}(r)$ across $[r_{\mathrm{hyp}}, r_{\mathrm{end}}]$; since $\theta_{\mathrm{phys}}(r) > \eta_2(r) - \pi$ for all $r$, crossing the lower nullcline $\theta = \eta_2(r) - \pi$ is impossible, contradicting $\theta_-(r_-) = \eta_2(r_-) - \pi$. Hence $\theta_-(r_{\mathrm{hyp}}) < \theta_{\mathrm{phys}}(r_{\mathrm{hyp}})$ and $\theta_-(r) < \theta_{\mathrm{phys}}(r) < 0$ across $[r_{\mathrm{hyp}}, r_-]$, proving both claims.
\end{proof}

\section{Contraction Bound and Proof of Proposition~\ref{prop:gsmb_contraction}}
\label{app:gsmb_proof}

Building on exponential dichotomy theory~\cite{Coppel1978}, the inclination ($\lambda$-) lemma~\cite{Palis1969}, and multiple-shooting conditioning~\cite{Keller1976}, Proposition~\ref{prop:gsmb_contraction} is stated formally under the sufficient analytical assumptions~(H1)--(H3)---which correspond to the practical runtime mechanisms of Sec.~II\,E without claiming unconditional equivalence for arbitrary potentials---and all three parts are proved together with the explicit bound on $C_{K}$.

\begin{proposition}[Physical-Ray Capture, Frame-Rotation Contraction Bound, and Backward Stability Inversion]
\label{prop:gsmb_contraction}
Consider the linear phase-space system~\eqref{eq:phase_ode} on a compact strictly hyperbolic forbidden domain $k^2(r; E) \le -k_{\min}^2 < 0$ partitioned by adaptive handoff radii $r_{\mathrm{hyp}} = x_{f,0} < x_{f,1} < \dots < x_{f,K} \le r_{\mathrm{end}}$, where $x_{f,k}^-$ and $x_{f,k}^+$ denote pre- and post-rescaling evaluation at $x_{f,k}$, respectively. Let $\phi_k \in (0,\pi)$ denote the angle in covariant eigenframe coordinates $\boldsymbol{\xi} = (c_s, c_u)^\top$ between the chord vector $\delta\boldsymbol{\xi}^{(k)} \equiv \boldsymbol{\xi}(\mathbf{x}_+^{(k)}) - \boldsymbol{\xi}(\mathbf{x}_-^{(k)})$ and the stable axis $(1,0)^\top$, let $\rho_k \equiv \min_{w \in [0,1]} \|\mathbf{x}^{(k)}(w; x_{f,k}^+)\|_2$ denote the minimum Euclidean norm along the normalized chord $\mathcal{C}_k$, and set $\kappa_k \equiv \kappa_{\mathbf{V}}(x_{f,k-1})\kappa_{\mathbf{V}}(x_{f,k})$. Assume:
\begin{enumerate}
\item[(H1)] \emph{(Initial straddling and sub-chord selection)} Each normalized chord $\mathcal{C}_k$ ($k \ge 0$) straddles {$\mathcal{E}^s(x_{f,k}) = \mathrm{span}\{\mathbf{x}_{\mathrm{phys}}(x_{f,k}; E)\}$,} with $\mathcal{C}_k = [\mathbf{x}_{m^*}^{(k)}, \mathbf{x}_{m^*+1}^{(k)}]$ for $k \ge 1$ selected across the sign change of the signed transverse coordinate $\ell_k(w) \equiv \langle \mathbf{J}\mathbf{x}_{\mathrm{phys}}(x_{f,k}; E),\, \mathbf{x}^{(k)}(w; x_{f,k}^-) \rangle$ (with $\mathbf{J} = \bigl(\begin{smallmatrix}0 & -1 \\ 1 & 0\end{smallmatrix}\bigr)$) and rescaled by $\gamma_k = \|\mathbf{x}_{m^*}^{(k-1)}(x_{f,k}^-)\|_2^{-1}$ so that $\|\mathbf{x}_-^{(k)}(x_{f,k}^+)\|_2 = 1$.
\item[(H2)] \emph{(Uniform chord transversality and non-degeneracy)} There exist constants $s_{\min} > 0$ and $\rho_{\min} > 0$ independent of $k$ such that $\sin\phi_k \ge s_{\min}$ and $\rho_k \ge \rho_{\min}$ for all $k \in \{0,\dots,K-1\}$ (automatically satisfied when $\|\mathbf{x}_+^{(k)} - \mathbf{x}_-^{(k)}\|_2 < 1$).
\item[(H3)] \emph{(Sub-horizon handoff spacing)} For a uniform contraction margin $\eta > 0$, the subdivision count $M \ge 3$ and handoff radii $\{x_{f,k}\}_{k=0}^K$ satisfy
\begin{equation}
\begin{split}
\int_{x_{f,k-1}}^{x_{f,k}} \Bigl[\Lambda_\theta(r) + 2\kappa_{\mathbf{V}}(r)&\bigl(|\theta_s'(r)| + |\theta_u'(r)|\bigr)\Bigr]\,dr \\
&\le \ln\!\left(\frac{\rho_{k-1}\sin\phi_{k-1}(M-1)}{\kappa_k}\right) - \eta
\end{split}
\label{eq:handoff_spacing_hyp}
\end{equation}
for all $k \in \{1,\dots,K\}$.
\end{enumerate}
Then:
\begin{enumerate}
\item \emph{(Physical-Ray Capture and Tilt Conditioning)} For every $k \ge 0$, a unique $w_k^* \in [0,1]$ satisfies $\mathbf{x}^{(k)}(w_k^*; x_{f,k}) \in \mathcal{E}^s(x_{f,k})$. Varying $\phi_k \in (0,\pi)$ rescales amplitude along $\mathbf{x}_{\mathrm{phys}}(r; E)$ in exact arithmetic, while $\sin\phi_k \ge s_{\min} > 0$ bounds the transverse conditioning factor $\csc\phi_k \le s_{\min}^{-1}$ in finite precision.
\item \emph{(Bounded Frame-Rotation Contraction)} Let $\Delta c_u^{(k)} \equiv |\langle \mathbf{n}_u(x_{f,k}),\, \mathbf{x}_+^{(k)} - \mathbf{x}_-^{(k)} \rangle|$ be the scale-normalized covariant unstable bracket width at generation $k$. Across $K$ generations,
\begin{equation}
\begin{split}
\Delta c_u^{(K)} &\le \frac{C_K\,\Delta c_u^{(0)}}{(M-1)^K}\exp\!\left(\int_{r_{\mathrm{hyp}}}^{x_{f,K}} \Lambda_\theta(r)\,dr\right) \\
&\le e^{-\eta K}\,\Delta c_u^{(0)},
\end{split}
\label{eq:contraction_bound}
\end{equation}
where $\Lambda_\theta(r) = \lambda_u(r; E) - \lambda_s(r; E) > 0$ [Eq.~\eqref{eq:spectral_gap}] and $C_K < \infty$ is bounded explicitly in Eq.~\eqref{eq:ck_bound}. For fixed $M$ with~\eqref{eq:handoff_spacing_hyp}, $\Delta c_u^{(K)}$ decays geometrically as $e^{-\eta K}$; for fixed partition $K$, increasing $M$ yields algebraic decay $\mathcal{O}\bigl((M-1)^{-K}\bigr)$.
\item \emph{(Retroactive Stability Inversion)} Integrating $\mathbf{x}_m^{(K)} \in \mathcal{C}_K$ backward in radius ($r: x_{f,K} \to r_{\min}$) inverts $\mathbf{A}(r; E) \mapsto -\mathbf{A}(r; E)$ and flips the saddle signs on $[r_{\mathrm{hyp}}, x_{f,K}]$ to $-\lambda_s(r; E) > 0$ and $-\lambda_u(r; E) < 0$, making the decaying physical ray attracting in reverse time and compressing the inner Pr\"ufer angle and slope error as $|\theta_K^{(m)}(r_{\min}) - \theta_{\mathrm{phys}}(r_{\min})| = \mathcal{O}(e^{-\eta K})$ and
\begin{equation}
\left|\tan\theta_K^{(m)}(r_{\min}) - \tan\theta_{\mathrm{exact}}(r_{\min})\right| = \mathcal{O}(e^{-\eta K})
\label{eq:backward_compression}
\end{equation}
(whenever $\cos\theta_{\mathrm{phys}}(r_{\min}) \neq 0$, with $\tan\theta_{\mathrm{exact}}(r_{\min}) \equiv \tan\theta_{\mathrm{phys}}(r_{\min})$).
\end{enumerate}
\end{proposition}

\noindent\emph{Proof overview.} The proof proceeds in three stages corresponding to a multi-step relay across $[r_{\mathrm{hyp}}, x_{f,K}]$: (1) linearity of the inter-fiber propagator preserves affine barycentric ordering across each handoff, guaranteeing a unique physical-ray crossing $w_k^* \in [0,1]$; (2) on each sub-interval $[x_{f,k-1}, x_{f,k}]$, Gronwall's inequality bounds the competition between exponential saddle stretching ($\Lambda_\theta$), eigenframe rotation leakage ($2\kappa_{\mathbf{V}}(|\theta_s'| + |\theta_u'|)$), and $(M-1)^{-1}$ barycentric subdivision, yielding geometric contraction $e^{-\eta K}$ under~\eqref{eq:handoff_spacing_hyp}; and (3) reversing the radial integration direction inverts the saddle eigenvalues ($\lambda_{s,u} \mapsto -\lambda_{s,u}$), converting the decaying physical orbit into an attracting solution that damps residual barrier errors during the inward sweep to $r_{\min}$.

\begin{proof}[Proof of Proposition~\ref{prop:gsmb_contraction}]
(1) Because $\mathcal{E}^s(r) = \mathrm{span}\{\mathbf{x}_{\mathrm{phys}}(r; E)\}$ is invariant under the orientation-preserving linear propagator $\mathcal{U}(r, x_{f,k})$ [Eqs.~\eqref{eq:liouville_det} and~\eqref{eq:phys_ray_invariance}], $\mathcal{C}_k$ maps onto an affine segment for $r \ge x_{f,k}$ preserving barycentric weights $w$, and positive rescaling $\gamma_{k+1}$ preserves both rays and weights. Hence the signed transverse coordinate $\ell_k(w)$ orthogonal to $\mathbf{x}_{\mathrm{phys}}(x_{f,k}; E)$ is affine on $[0,1]$ with $\ell_k(0)\ell_k(1) \le 0$ and nonzero slope (since $\sin\phi_k \ge s_{\min} > 0 \implies \Delta c_u^{(k)} > 0$), giving a unique root $w_k^* \in [0,1]$. Expanding $\mathbf{x}_+^{(k)} - \mathbf{x}_-^{(k)} = \alpha_s\mathbf{e}_s(x_{f,k}) + \alpha_u\mathbf{e}_u(x_{f,k})$ and pairing with $\mathbf{n}_u(x_{f,k})$ [Eq.~\eqref{eq:covariant_eigenvectors}] gives $\Delta c_u^{(k)} = |\alpha_u|$, while $\sin\phi_k \ge s_{\min} > 0$ bounds $\csc\phi_k \le s_{\min}^{-1}$ in floating-point arithmetic (and forward hyperbolic growth of $\delta c_u$ relative to $\delta c_s$ tilts propagated chords away from $\phi_k \to 0$).

(2) From $\|\mathbf{x}_-^{(k-1)}(x_{f,k-1}^+)\|_2 = 1$ and the chord parameterization $\mathbf{x}^{(k-1)}(w; x_{f,k-1}^+)$, the reverse triangle inequality gives for all $w \in [0,1]$
\begin{equation}
\begin{split}
\|\mathbf{x}^{(k-1)}(w; x_{f,k-1}^+)\|_2 &\ge 1 - w\bigl\|\mathbf{x}_+^{(k-1)}(x_{f,k-1}^+) - \mathbf{x}_-^{(k-1)}(x_{f,k-1}^+)\bigr\|_2 \\
&\ge 1 - \frac{\|\mathbf{V}(x_{f,k-1})\|_2}{\sin\phi_{k-1}}\,\Delta c_u^{(k-1)},
\end{split}
\label{eq:chord_norm_floor}
\end{equation}
so $\rho_{k-1} > 0$ whenever the normalized chord length is less than unity.

On $[x_{f,k-1}, x_{f,k}]$ ($k = 1,\dots,K$), the covariant difference $\delta\boldsymbol{\xi}(r) \equiv \boldsymbol{\xi}_+^{(k-1)}(r) - \boldsymbol{\xi}_-^{(k-1)}(r)$ and selected trajectory $\boldsymbol{\xi}_{m^*}(r) \equiv \boldsymbol{\xi}(\mathbf{x}_{m^*}^{(k-1)}(r))$ satisfy $d\|\delta\boldsymbol{\xi}\|_2/dr \le \bigl(\lambda_u(r) + \|\mathbf{M}_{\mathrm{rot}}(r)\|_2\bigr)\|\delta\boldsymbol{\xi}\|_2$ and $d\|\boldsymbol{\xi}_{m^*}\|_2/dr \ge \bigl(\lambda_s(r) - \|\mathbf{M}_{\mathrm{rot}}(r)\|_2\bigr)\|\boldsymbol{\xi}_{m^*}\|_2$ [Eq.~\eqref{eq:rotating_frame_ode}]. Applying Gronwall's inequality with~\eqref{eq:rotation_norm_bound}, $\|\delta\boldsymbol{\xi}(x_{f,k-1}^+)\|_2 = \csc\phi_{k-1}\,\Delta c_u^{(k-1)}$, $\|\boldsymbol{\xi}_{m^*}(x_{f,k-1}^+)\|_2 \ge \rho_{k-1}\|\mathbf{V}(x_{f,k-1})\|_2^{-1}$, rescaling by $\gamma_k = \|\mathbf{x}_{m^*}^{(k-1)}(x_{f,k}^-)\|_2^{-1}$, and applying $(M-1)^{-1}$ barycentric subdivision with $|\delta c_u(x_{f,k}^-)| \le \|\delta\boldsymbol{\xi}(x_{f,k}^-)\|_2$ yields
\begin{equation}
\begin{split}
\Delta c_u^{(k)} &\le \frac{\kappa_k\,\Delta c_u^{(k-1)}}{\rho_{k-1}\sin\phi_{k-1}(M-1)} \\
&\quad\times \exp\!\left(\int_{x_{f,k-1}}^{x_{f,k}} \Bigl[\Lambda_\theta(r) + 2\kappa_{\mathbf{V}}(r)\bigl(|\theta_s'(r)| + |\theta_u'(r)|\bigr)\Bigr]\,dr\right).
\end{split}
\label{eq:single_step_contraction}
\end{equation}
Since $\Delta\theta(r) \in (0, \pi)$ is bounded away from $0$ and $\pi$ on the compact strictly hyperbolic interval $[r_{\mathrm{hyp}}, r_{\mathrm{end}}]$, we have $\kappa_{\max} \equiv \sup_{r \in [r_{\mathrm{hyp}}, r_{\mathrm{end}}]} \kappa_{\mathbf{V}}(r) < \infty$ and $\kappa_k \le \kappa_{\max}^2$. Multiplying~\eqref{eq:single_step_contraction} over $k = 1,\dots,K$ gives the first bound in~\eqref{eq:contraction_bound} with
\begin{equation}
\begin{split}
C_K &\le \left(\prod_{k=1}^{K}\frac{\kappa_k}{\rho_{k-1}\sin\phi_{k-1}}\right) e^{2\Theta_{\mathrm{rot}}} \\
&\le \left(\frac{\kappa_{\max}^2}{\rho_{\min}s_{\min}}\right)^K e^{2\Theta_{\mathrm{rot}}} < \infty,
\end{split}
\label{eq:ck_bound}
\end{equation}
where $\Theta_{\mathrm{rot}} \equiv \int_{r_{\mathrm{hyp}}}^{x_{f,K}} \kappa_{\mathbf{V}}(r)\bigl[|\theta_s'(r)| + |\theta_u'(r)|\bigr]\,dr < \infty$. Under $\sin\phi_{k-1} \ge s_{\min} > 0$, $\rho_{k-1} \ge \rho_{\min} > 0$, and~(H3) [Eq.~\eqref{eq:handoff_spacing_hyp}], each generation contracts by at least $e^{-\eta} < 1$, giving $\Delta c_u^{(K)} \le e^{-\eta K}\,\Delta c_u^{(0)}$.

(3) Setting $\tilde{r} = x_{f,K} - r$ transforms~\eqref{eq:phase_ode} into $d\mathbf{x}/d\tilde{r} = -\mathbf{A}(r; E)\mathbf{x}$, reversing the eigenvalue signs on $[r_{\mathrm{hyp}}, x_{f,K}]$ to $\lambda_s^{\mathrm{back}} = -\lambda_s > 0$ and $\lambda_u^{\mathrm{back}} = -\lambda_u < 0$ [Eq.~\eqref{eq:eigenvalues}] and swapping angular stability and instability [Eq.~\eqref{eq:angular_stability}]. Expanding $\mathbf{x}_m^{(K)}(x_{f,K}) = \beta_s\mathbf{x}_{\mathrm{phys}}(x_{f,K}; E) + \beta_u\mathbf{x}_{\mathrm{div}}(x_{f,K}; E)$ (where $\mathbf{x}_{\mathrm{div}}$ spans a complementary unstable subspace) with $|\beta_u/\beta_s| = \mathcal{O}(e^{-\eta K})$ by part~(2) and integrating backward to $r_{\min}$ (with periodic unit-circle rescaling of $R(r)$) damps the divergent mode across the barrier by $\exp\!\bigl(-\int_{r_{\mathrm{hyp}}}^{x_{f,K}} \Lambda_\theta(r)\,dr\bigr) \ll 1$ relative to $\mathbf{x}_{\mathrm{phys}}$ under $\|\mathbf{M}_{\mathrm{rot}}(r)\|_2 \ll \Lambda_\theta(r)$ while keeping $[r_{\min}, r_{t,1}]$ well conditioned, yielding~\eqref{eq:backward_compression}.
\end{proof}

\bibliographystyle{apsrev4-2}
\bibliography{references}

@article{Numerov1924,
  author  = {Numerov, B. V.},
  title   = {A method of extrapolation of perturbations},
  journal = {Mon. Not. R. Astron. Soc.},
  volume  = {84},
  pages   = {592},
  year    = {1924}
}

@article{Cooley1961,
  author  = {Cooley, J. W.},
  title   = {An improved eigenvalue corrector formula for solving the {Schr\"odinger} equation for central fields},
  journal = {Math. Comput.},
  volume  = {15},
  pages   = {363},
  year    = {1961}
}

@article{Johnson1977,
  author  = {Johnson, B. R.},
  title   = {New numerical methods applied to solving the one-dimensional eigenvalue problem},
  journal = {J. Chem. Phys.},
  volume  = {67},
  pages   = {4086},
  year    = {1977}
}

@article{Manolopoulos1986,
  author  = {Manolopoulos, D. E.},
  title   = {An improved log derivative method for inelastic scattering},
  journal = {J. Chem. Phys.},
  volume  = {85},
  pages   = {6425},
  year    = {1986}
}

@article{Aguilar1971,
  author  = {Aguilar, J. and Combes, J. M.},
  title   = {A class of analytic perturbations for one-body {Schr\"odinger} Hamiltonians},
  journal = {Commun. Math. Phys.},
  volume  = {22},
  pages   = {269},
  year    = {1971}
}

@article{Balslev1971,
  author  = {Balslev, E. and Combes, J. M.},
  title   = {Spectral properties of many-body {Schr\"odinger} operators with dilatation-analytic interactions},
  journal = {Commun. Math. Phys.},
  volume  = {22},
  pages   = {280},
  year    = {1971}
}

@book{Moiseyev2011,
  author    = {Moiseyev, N.},
  title     = {Non-Hermitian Quantum Mechanics},
  publisher = {Cambridge University Press},
  address   = {Cambridge},
  year      = {2011}
}

@article{Seideman1992,
  author  = {Seideman, T. and Miller, W. H.},
  title   = {Calculation of the cumulative reaction probability via a discrete variable representation with absorbing boundary conditions},
  journal = {J. Chem. Phys.},
  volume  = {96},
  pages   = {4412},
  year    = {1992}
}

@article{Riss1993,
  author  = {Riss, U. V. and Meyer, H.-D.},
  title   = {Calculation of resonance energies and widths using the complex absorbing potential method},
  journal = {J. Phys. B: At. Mol. Opt. Phys.},
  volume  = {26},
  pages   = {4503},
  year    = {1993}
}

@article{Kato1957,
  author  = {Kato, T.},
  title   = {On the eigenfunctions of many-particle systems in quantum mechanics},
  journal = {Commun. Pure Appl. Math.},
  volume  = {10},
  pages   = {151},
  year    = {1957}
}

@article{Molano2021,
  author  = {Molano, J. S. and Arango, C. A.},
  title   = {A mathematical phase-space representation of the 1-dimensional {Schr\"odinger} equation},
  journal = {Chem. Phys. Lett.},
  volume  = {762},
  pages   = {138171},
  year    = {2021}
}

@misc{Bandyopadhyay2025,
  author        = {Bandyopadhyay, Shalmali and \c{C}etinkaya, F. Ay\c{c}a and Cuchta, Tom},
  title         = {Pr\"ufer Transformation and Spectral Analysis for a {S}turm--{L}iouville-Type Equation},
  year          = {2025},
  eprint        = {2508.12369},
  archivePrefix = {arXiv},
  primaryClass  = {math.CA}
}

@article{Solarte2025,
  author  = {Solarte, I. E. and Barreiro-Medina, S. and Arango, C. A.},
  title   = {Phase-space Propagation of the Time-independent {Schr\"odinger} Equation for One-dimensional Rectangular Potential Systems},
  journal = {Int. J. Theor. Phys.},
  volume  = {64},
  pages   = {208},
  year    = {2025},
  doi     = {10.1007/s10773-025-06064-9}
}

@article{Wigner1955,
  author  = {Wigner, E. P.},
  title   = {Lower limit for the energy derivative of the scattering phase shift},
  journal = {Phys. Rev.},
  volume  = {98},
  pages   = {145},
  year    = {1955}
}

@article{Smith1960,
  author  = {Smith, F. T.},
  title   = {Lifetime matrix in collision theory},
  journal = {Phys. Rev.},
  volume  = {118},
  pages   = {349},
  year    = {1960}
}

@book{Coppel1978,
  author    = {Coppel, W. A.},
  title     = {Dichotomies in Stability Theory},
  series    = {Lecture Notes in Mathematics},
  volume    = {629},
  publisher = {Springer-Verlag},
  address   = {Berlin},
  year      = {1978}
}

@article{Palis1969,
  author  = {Palis, J.},
  title   = {On {Morse--Smale} dynamical systems},
  journal = {Topology},
  volume  = {8},
  pages   = {385},
  year    = {1969}
}

@book{Keller1976,
  author    = {Keller, H. B.},
  title     = {Numerical Solution of Two Point Boundary Value Problems},
  series    = {CBMS-NSF Regional Conference Series in Applied Mathematics},
  volume    = {24},
  publisher = {SIAM},
  address   = {Philadelphia},
  year      = {1976}
}

@book{MorseFeshbach1953,
  author    = {Morse, Philip M. and Feshbach, Herman},
  title     = {Methods of Theoretical Physics, Part I},
  publisher = {McGraw-Hill},
  address   = {New York},
  year      = {1953},
  note      = {Sec.~1.3}
}

@book{Horn2012,
  author    = {Horn, Roger A. and Johnson, Charles R.},
  title     = {Matrix Analysis},
  edition   = {2nd},
  publisher = {Cambridge University Press},
  address   = {Cambridge},
  year      = {2012},
  note      = {Sec.~1.4}
}

@article{Prufer1926,
  author  = {Pr{\"u}fer, H.},
  title   = {Neue Herleitung der {S}turm--{L}iouville'schen Reihenentwicklung stetiger Funktionen},
  journal = {Math. Ann.},
  volume  = {95},
  pages   = {499--518},
  year    = {1926}
}

@book{Atkinson1964,
  author    = {Atkinson, F. V.},
  title     = {Discrete and Continuous Boundary Problems},
  publisher = {Academic Press},
  address   = {New York},
  year      = {1964}
}

@book{Zettl2005,
  author    = {Zettl, A.},
  title     = {Sturm--Liouville Theory},
  publisher = {American Mathematical Society},
  address   = {Providence, RI},
  year      = {2005}
}

@article{Johnson1978,
  author  = {Johnson, B. R.},
  title   = {The renormalized {N}umerov method applied to calculating bound states of the coupled-channel {S}chr{\"o}dinger equation},
  journal = {J. Chem. Phys.},
  volume  = {69},
  pages   = {4678--4688},
  year    = {1978}
}

@article{Magnus1954,
  author  = {Magnus, Wilhelm},
  title   = {On the exponential solution of differential equations for a linear operator},
  journal = {Commun. Pure Appl. Math.},
  volume  = {7},
  pages   = {649--673},
  year    = {1954}
}

@article{Pechukas1966,
  author  = {Pechukas, Philip and Light, John C.},
  title   = {On the Exponential Form of Time-Displacement Operators in Quantum Mechanics},
  journal = {J. Chem. Phys.},
  volume  = {44},
  pages   = {3897--3912},
  year    = {1966}
}

@article{Light1976,
  author  = {Light, John C. and Walker, Robert B.},
  title   = {An {R}-matrix approach to the solution of coupled equations for atom--molecule reactive scattering},
  journal = {J. Chem. Phys.},
  volume  = {65},
  pages   = {4272--4282},
  year    = {1976}
}

@article{Blanes2009,
  author  = {Blanes, S. and Casas, F. and Oteo, J. A. and Ros, J.},
  title   = {The {M}agnus expansion and some of its applications},
  journal = {Phys. Rep.},
  volume  = {470},
  pages   = {151--238},
  year    = {2009}
}

@article{Colbert1992,
  author  = {Colbert, Daniel T. and Miller, William H.},
  title   = {A novel discrete variable representation for quantum mechanical reactive scattering via the {S}-matrix {K}ohn method},
  journal = {J. Chem. Phys.},
  volume  = {96},
  number  = {3},
  pages   = {1982--1991},
  year    = {1992},
  doi     = {10.1063/1.462100}
}
\end{document}